\documentclass[]{vgtc}
\usepackage{stfloats}
\ifpdf%                                % if we use pdflatex
  \pdfoutput=1\relax                   % create PDFs from pdfLaTeX
  \usepackage{graphicx}                % allow us to embed graphics files
  \DeclareGraphicsExtensions{.pdf,.png,.jpg,.jpeg} % for pdflatex we expect .pdf, .png, or .jpg files
\else%                                 % else we use pure latex
  \usepackage{graphicx}                % allow us to embed graphics files
  \DeclareGraphicsExtensions{.eps}     % for pure latex we expect eps files
\fi%

\graphicspath{{figures/}{pictures/}{images/}{./}} % where to search for the images

\usepackage{microtype}                 % use micro-typography (slightly more compact, better to read)
\PassOptionsToPackage{warn}{textcomp}  % to address font issues with \textrightarrow
\usepackage{textcomp}                  % use better special symbols
\usepackage{mathptmx}                  % use matching math font
\usepackage{times}                     % we use Times as the main font
\usepackage{cite}                      % needed to automatically sort the references
\usepackage{tabu}                      % only used for the table example
\usepackage{booktabs}    
\usepackage{paralist}

\usepackage{xcolor}
\usepackage{hyperref}
\usepackage{paralist}
\onlineid{1897}

\vgtccategory{Research}

\vgtcinsertpkg

\preprinttext{To appear in an IEEE VGTC sponsored conference.}

\title{Using a virtual reality interview simulator to explore factors influencing people’s behavior}

\author{Xinyi Luo\thanks{e-mail: 2020090903013@std.uestc.edu.cn}\\ %
     \parbox{1.45in}{\scriptsize \centering University of Electronic Science and Technology of China } %
\and Yuyang Wang\thanks{e-mail: yuyangwang@ust.hk}\\ %
     \parbox{1.45in}{\scriptsize \centering Hong Kong University of Science and Technology (Guangzhou) }
\and Lik-Hang Lee\thanks{e-mail: lhleeac@connect.ust.hk}\\ %
     \parbox{1.45in}{\scriptsize \centering The Hong Kong Polytechnic University}%
\and Zihan Xing\\ %
     \parbox{1.45in}{\scriptsize \centering Beijing Normal University Hong Kong Baptist University United International College }%
\and Shan Jin\\ %
     \parbox{1.45in}{\scriptsize \centering Hong Kong University of Science and Technology (Guangzhou) }%
\and Boya Dong\\ %
     \parbox{1.45in}{\scriptsize \centering South China University of Technology }%
\and Yuanyi Hu\\ %
     \parbox{1.45in}{\scriptsize \centering Guangdong Medical University }%
\and Zeming Chen\\ %
     \parbox{1.45in}{\scriptsize \centering South China University of Technology }%
\and Jing Yan\\ %
     \parbox{1.45in}{\scriptsize \centering Taiyuan University of Technology }%
\and Pan Hui\\ %
     \parbox{1.45in}{\scriptsize \centering Hong Kong University of Science and Technology (Guangzhou)}     
     }

\abstract{
Virtual reality interview simulator (VRIS) provides an effective and manageable approach for candidates prone to being very nervous during interviews, yet, the major anxiety-inducing elements remain unknown. During an interview, the anxiety levels, overall experience, and performance of interviewees might be affected by various circumstances. By analyzing electrodermal activity and questionnaire, we investigated the influence of five variables: (I) \textit{Realism}; (II) \textit{Question type}; (III) \textit{Interviewer attitude}; (IV) \textit {Timing}; and (V) \textit{Preparation}. As such, an orthogonal design $L_8(4^1 \times 2^4)$ with eight experiments ($O A_8$ matrix) was implemented, in which 19 
college students took part in the experiments. Considering the anxiety, overall experience, and performance of the interviewees, results indicate that \textit{Question type} plays a major role; secondly, \textit{Realism}, \textit{Preparation}, and \textit{Interviewer attitude} all have some degree of influence; lastly, \textit {Timing} have little to no impact. Specifically, professional interview questions elicited a greater degree of anxiety than personal ones among the categories of interview questions. This work contributes to our understanding of anxiety-stimulating factors during job interviews in virtual reality and provides cues for designing future VRIS.
} % end of abstract

\CCScatlist{ 
  \CCScat{I.3.7}{Three-Dimensional Graphics and Realism}%
{Virtual reality}{};
  \CCScat{H.5.2}{User Interfaces}{Training, help, and documentation}{}
}
\begin{document}



\firstsection{Introduction}

\maketitle

The global pandemic in recent years has aggravated the situation where many companies are hiring fewer employees, which has led to a series of anxieties among college students~\cite{mok2021covid}, one of which is interview anxiety disorder (IAD)-anxiety manifests itself in the form of speech disturbances, socially inappropriate behaviors as well as other nervous jitters~\cite{levine2002women}. Furthermore, past research has shown that as interview anxiety increases, people tend to develop more protective self-presentational tactics, making it difficult for interviewees to perform well and decreasing competitiveness~\cite{feiler2016behavioral}.
%\IEEEpubidadjcol
Virtual reality has become a common tool in the therapeutic field, such as substance~\cite{albright2021innovative}, high-functioning autism~\cite{kandalaft2013virtual, didehbani2016virtual} and eating disorders~\cite{clus2018use}. Virtual reality exposure therapy (VRET) is an important treatment for various anxiety disorders. By mimicking social circumstances for social distress patients~\cite{emmelkamp2020virtual} and using computer-generated virtual scenes~\cite{kampmann2016meta}, patients can be exposed to an environment with virtual social situations and interactions that target diverse social fears in a controllable way. For example, one study indicated the utility of VR in inducing stressful reactions through a combination of stressors for healthy subjects \cite{kerous2020examination}.
However, the factors that actually influence an interviewee's anxiety, overall experience, and interview performance in a VRET interview are still unclear and poorly understood, especially when multiple factors are assembled in one system; thus, our research serves as a pioneer for such an investigation. 

So far, researchers have also conducted extensive research on systems for VR interview training and evaluation.
For example, a virtual job interviewing practice system has been designed for high-anxiety populations like people with Autism Spectrum Disorder and former convicts~\cite{hartholt2019virtual}, a VR–based job interview training platform has been developed for autistic individuals to practice interviewing skills in a less anxiety-inducing virtual context~\cite{adiani2022career}, an agent-based VR training and multidimensional evaluation system has been created for introverted college students to cope with interview anxiety~\cite{jin2019developing}, a job training simulation environment has been presented for young people who are out of employment, education, or training with social cue recognition techniques~\cite{gebhard2014exploring}.

Nevertheless, most studies today validate and evaluate an entire product without examining the multiple factors separately. Not only do we know very little about the significant elements that truly influence interview anxiety and the overall experience (e.g., cognitive load, discomfort~\cite{Wang2021cog,Wang2021speed}),
but we also need a solid understanding of how each factor affects the interviewee's external performance (e.g., verbal expressions, eye contact, body movements).
For graduates to be competitive in the job market and for employers to pick prospective workers, they must demonstrate strong performance during job interviews and have practical anxiety management abilities. Therefore, a study on factors contributing to people's interview anxiety is required to create successful training programs and develop tailored therapy approaches.

In terms of anxiety, previous studies have found that visual display \cite{kwon2013level}, interview questions \cite{hartwell2019we}, interviewer's attitude \cite{kwon2009study}, timing \cite{schwartz2015comparison} and preparation \cite{gantt2013effect} can all have an impact on interviewee's anxiety. 
In addition to feelings of anxiety, other studies have investigated the impact of virtual reality on overall experience (e.g., discomfort, eyestrain, psychosocial stress).
Some studies have found that apparatus has an impact on quality of experience, with HMDs experiencing higher eyestrain and visual discomfort than PCs \cite{souchet2022short}; some have found the V-TSST (virtual environment versions of the TSST) is effective at inducing psychosocial stress which can lead to poor physical and psychological health outcomes, though the magnitude of this response is less than the traditional TSST \cite{helminen2019meta}; Kothgassner et al. have indicated that the perceived social presence did not differ over time in the VR TSST conditions as the main effect of time was not significant \cite{kothgassner2021habituation}.

Meanwhile, interviewees' performances have also been proven to be influenced by factors within a virtual environment. For example, prior studies have identified the relationship between the number of completed virtual interviews and improved interviewing skills or performance as the mechanism for getting a job offer \cite{smith2017mechanism}; the results of M Barreda-Ángeles et al. have shown that, compared to the neutral audience, the negative audience elicited increases in skin conductance level and heart rate variability, decreases in voice intensity, and a higher ratio of silent parts in the speech, as well as a more negative self-reported valence, higher anxiety, and lower social presence \cite{barreda2020users}.

As such, we developed the VRIS, where the above factors were introduced and investigated within an orthogonal design to examine their significance separately. 

In this paper, we question whether the above five factors (I) \textsc{Realism}; (II) \textsc{Question type}; (III) \textsc{Interviewer attitude}; (IV) \textsc {Timing}; and (V) \textsc{Preparation} will indeed significantly influence interviewee's anxiety during a job interview. We hypothesized that all of these factors could potentially have a pronounced effect on interview anxiety.
% 3. our main contribution - as the RQs
% 4. paper structure.
Consider the above mentioned, the current study proposed five research questions (RQs) with the following hypothesis (H):

\begin{enumerate}
    \item RQ1: How do different interview questions affect the interviewee?
    
     \textit{H1: }\textit{Professional interview questions can cause more anxiety, worse overall experience, and poorer performance than personal questions.}
    \item RQ2: How does preparation for an interview affect the interviewee?

     \textit{H2: }\textit{Being unprepared for an interview can be more nervous and uncomfortable than being well prepared for the content of the interview.}
    \item RQ3: How does the timing of the answers to interview questions affect the interviewee?

     \textit{H3: }\textit{Timed answers can be more nerve-wracking than untimed answers, leading to worse performance.}
    \item RQ4: How do different levels of realism affect the interviewee?
    
    \textit{H4: }\textit{A more realistic scenario would make interviewees more nervous and reduce their eye contact with the interviewer.}
    \item RQ5: How does the interviewer's attitude affect the interviewee?

    \textit{H5: }\textit{Compared to an interviewer with a positive attitude, an interviewer with a negative attitude will elicit increases in skin conductance response, decreases in eye contact, a higher cognitive load during the interview, as well as more unsatisfactory performance.}

    % \item RQ6: How do the above five factors affect the interviewer's anxiety level in order of importance?
    % \item \textsc{Hypothesis 6}: We suppose that the importance of factors in descending order as follows: \textsc{Interviewer attitide} $>$ \textsc{Level of Realism} $>$ \textsc{Type of Interview Questions} $>$ \textsc{With or Without Preparation} $>$ \textsc {Timed or Untimed Answers}.
\end{enumerate}

We then conducted an orthogonal experiment design with eight different interview conditions using a mixed level $L_8(4^1 \times 2^4) $ orthogonal table including all five factors above to examine the significance of each factor on interview anxiety, overall experience, and performance. In addition, we measured the interviewee's electrodermal activity(EDA) during the interview, given that increased EDA has been associated with anxiety. Finally, in all eight conditions, we asked the interviewee to fill out a self-rated anxiety questionnaire and NASA-TLX criteria once the interview was completed. The interviewer would also rate the interviewee's performance during the interview. 
Followed by the mixed-effects model and the associated post-hoc analysis for the data collected from questionnaires and electrodermal activity, we found that all five of the above factors had different levels of influence on interviewees' anxiety, overall experience and interview performance, among which \textsc{Type of Interview Questions} had the most significant impact, in particular, the professional questions significantly increased the interviewee's anxiety, discomfort, electrodermal activity and cognitive workload.

The proposed work aims to identify anxiety-stimulating factors during job interviews in virtual reality and provides detailed insights into designing future VRIS.

\section{Related Work}

Relevant prior work includes studies of psychotherapy, online interview systems, and virtual reality interview training. This section summarizes those works separately. 

% VR在心理治疗应用
\subsection{Psychotherapy in Virtual Reality}

Since the inception of virtual reality, several psychotherapy pieces of research have been undertaken in virtual settings, and the idea of employing them to treat psychiatric illnesses has been investigated.
% 首次对社交恐惧症进行了VRET研究
As social concern about social phobia gradually increased, North et al. first used virtual reality exposure therapy for social phobia: they developed a virtual auditorium that could be triggered in real-time~\cite{north1998virtual}. 
The audience and audio clips would respond to the experimenter's voice in the auditorium, prompting the experimenter to speak louder and more loudly.
Based on feedback from the questionnaire results, the experiment demonstrated that Virtual Reality Exposure Therapy (VRET) effectively mitigated anxiety symptoms during presentations. Although the feedback from the experiment mainly was auditory information, it highlights the need for further research on the relationship between VRET and human psychology.

% VRET虚拟现实暴露疗法
Further studies have implied that there are three preconditions for the treatment of anxiety disorders through VRET, including immersion, anxiety, and presence~\cite{huang2021application}. 
Parsons et al. reported 21 case studies~\cite{parsons2008affective} confirming that VRET can effectively treat arachnophobia (Garcia et al.~\cite{garcia2002virtual}), flying phobia (Banos and Botella~\cite{banos2002virtual}), phobia of public places ( Botella et al.~\cite{ botella2007virtual}),
acrophobia(Coelho et al.~\cite{coelho2006virtual})

% VRET在ASD儿童社会认知训练
%Some studies have focused on providing social cognitive interaction training to children with autism, 
Some researchers have emphasized social cognition interaction training for autistic youngsters.
Didehbani et al. designed a virtual reality social cognitive training to improve the social skills of children with ASD~\cite{didehbani2016virtual}. %They measured emotion recognition, social attribution, attention, and executive functioning, and the study's results confirmed that the virtual reality platform effectively improved the social impairments common in ASD.
The research findings, which tested emotion detection, social attribution, attention, and executive functioning, revealed that the virtual reality platform successfully ameliorated the social deficits typical of ASD.

% VRET对发育障碍年轻人社交能力提升
In addition to virtual reality social cognitive training for children with ASD, Burke et al. designed Virtual Interactive Training Agents (ViTA) to develop social skills and reduce anxiety in young people with ASD and other developmental disorders \cite{burke2018using}.  
%Studies have shown that experimenters with ViTA training have a practical improvement in identifying strengths, self-promote, self-advocate, and answering situational questions.
According to studies, experimenters who have received ViTA training are much better at recognizing their abilities, promoting themselves, self-promoting, self-advocating, and responding to situational queries.
%modified the below when appropriate
In contrast, our study relates virtual reality to job interviews, the most prevalent situation in business practice. In the era of the metaverse, we anticipate migrating job interviews to the virtual realm. At the same time, our research exposes user behaviors in the settings of psychological factors and virtual job interviews.

% VR与线上面试系统
\subsection{Virtual Reality and Online Interview Systems}

% 线上视频面试与人工智能系统
%Due to the COVID-19 epidemic, recruitment formats are gradually changing, with more and more organizations using online video interviews in their hiring process. 
Due to the global pandemic and rising economic costs, firms are progressively incorporating online video interviews into their hiring and appointment procedures. In addition, the rapid development of artificial intelligence (AI) has driven the applications of automatic scoring, where AI interview systems can score interviewees' hard and soft skills based on the content of their answers, facial movements, eye contact, and speaking tones in the video~\cite{harwell2019face}, which successfully increase the effectiveness of interviews and assist professionals in assessing their candidates more quickly.

% 计算机模拟提高面试心理准备
Online interview formats have driven the development of online interview simulation training. For example, Aysina et al. created Job Interview Simulation Training (JIST) to improve psychological preparation for job interviews among the pre-retirement unemployed~\cite{aysina2017using}.
The experiment showed that having interviewees practice interviews over and over in a stress-free environment made them much more psychologically ready for the actual interview(s), which could help demonstrate the relationship between JIST and increased re-employment among pre-retirement job seekers in the future.

% VR在线上面试系统的有效性
Other studies have found that highly interactive virtual reality role-play training based on behavioral learning principles is more effective than traditional role-play training in training other types of interpersonal skills. 
Smith et al. developed a study to test a role-play simulation ``virtual reality job interview training'' (VR-JIT) for the feasibility and effectiveness of improving job-related interview content and interviewees' performance-related interview skills in individuals with ASD \cite{smith2014virtual}. Their findings demonstrate that VR-JIT can improve job interview skills in individuals with ASD.
Among current online interview systems, our research investigates the viability of conducting job interview preparations in immersive settings. Additionally, we specifically focus on the effect of different factors on interviewees' anxiety levels.

\section{VRIS Framework}

\subsection{Orthogonal experimental design}
The orthogonal experimental design is an efficient method to study the effect of multiple factors within one system simultaneously compared to the conventional methods of studying each factor separately by selecting one of the variables to change its parameters and fixing the rest of the variables. It selects some representative combinations from a full-scale test according to the modern algebra of Galois theory \cite{addelman1962orthogonal,huffcutt2011understanding}. The orthogonal table based on orthogonality ensures that the effects of all factors are obtained with a minimum number of trials.

We identified five independent variables that potentially affect interviewee's anxiety levels during a job interview in virtual reality: (I) \textsc{Level of Realism}
(4 levels); (II) \textsc{Type of Interview Questions}
(2 levels); (III) \textsc{Interviewer attitide}
(2 levels); (IV) \textsc {Timed or Untimed Answers}
(2 levels); and (V) \textsc{With or Without Preparation}
(2 levels). In order to determine the relative importance of these five variables and find out what factor most stimulates the interviewee's anxiety, we constructed an orthogonal fractional factorial design to arrange the tests by using a mixed level  $L_8(4^1 \times 2^4) $ orthogonal table with all these five variables in a total of eight sets of conditions, as shown in \autoref{tab:experiment}. 
For example, to conduct the seventh experimental group, the participants had to test with the Oculus Quest 2 HMD in a virtual job interview environment and be asked ten professional questions without preparation before the interview began. Each question will be timed 30 seconds for the answer by an interviewer with a negative attitude (e.g., passive body movements and negative verbal feedback).

\begin{figure}[ht]
\centering
\includegraphics[width=0.9\linewidth]{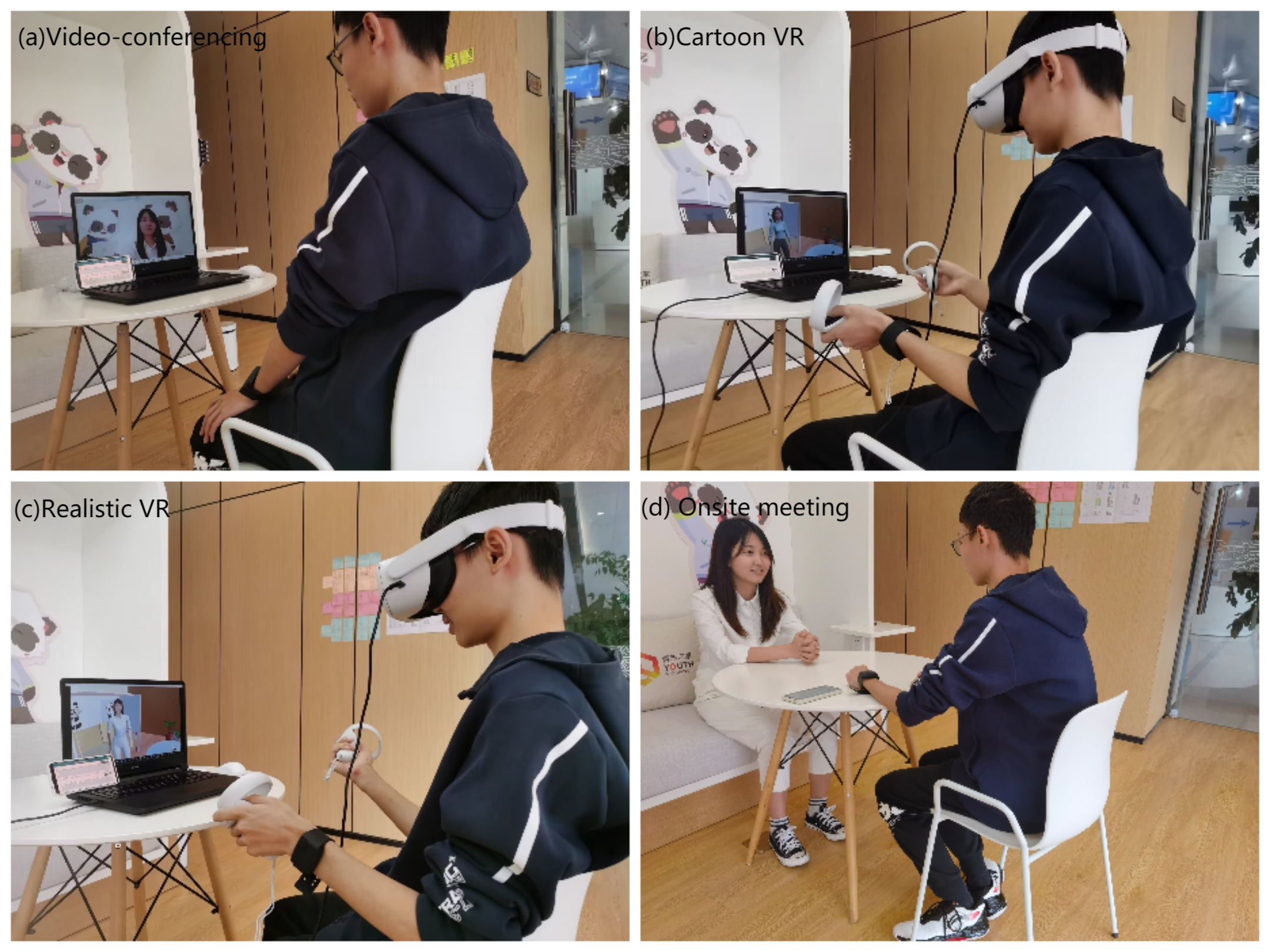}
\caption{Four experiment settings for job interview simulation under the conditions of (a) video-conferencing(\textit{PC}), (b) cartoon VR (\textit{VR1};), (c) realistic VR (\textit{VR2};), (d) onsite meeting (\textit{REAL}).}
\label{fig:interview}
\end{figure}

\begin{figure}[ht]
\centering
\includegraphics[width=0.9\linewidth]{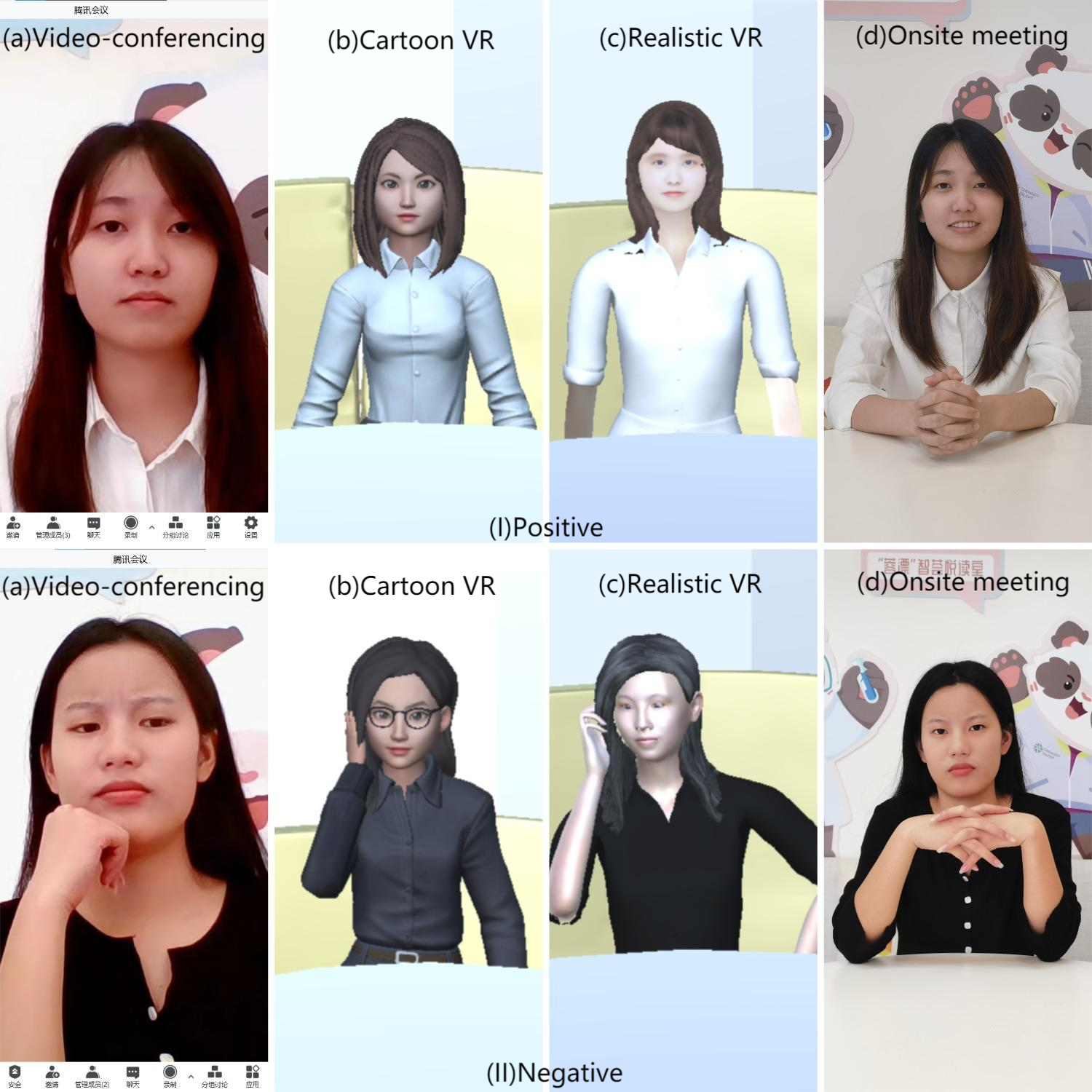}
\caption{Four levels of realism from the least to most immersive(from left to right): video-conferencing(\textit{VR1}), cartoon VR (\textit{VR1};), realistic VR (\textit{VR2};), onsite meeting (\textit{REAL}). Two types of attitude: positive and negative(from top to bottom).}
\label{fig:realism}
\end{figure}

The first independent variable (I) \textsc{Level of Realism} corresponded to the visual display of the interviewer. We were aware that a video conference(e.g., through Skype, Tencent Meeting) would be considered quite "real" as an online meeting using video conferencing is generally a well-accepted form of a remote interview. In our context, "realism" refers to the level of immersion. Previous studies have found that higher visual display levels provoke more anxiety and the sense of presence \cite{kwon2013level}. Therefore, in order to create four different kinds of realism, we set four conditions representing a continuous spectrum of immersiveness, i.e., the least to the most immersive, including an interviewer presented by video conference on a laptop computer (\textit{PC}), a low-poly and cartoon-like 3D avatar representing the human interviewer (\textit{VR1}), a realistic interviewer with a high fidelity 3D human avatar (\textit{VR2}) and a face-to-face real human interviewer (\textit{REAL}) (see \autoref{fig:realism}). 
In the experiments under \textit{PC} condition, interviewees conducted video interviews with live interviewers via video conference. In the experiments under \textit{VR1} and \textit{VR2} condition, animated sequences of the interviewer's avatar were played automatically by the software in a VR headset, and a conversation with a live interviewer was conducted via voice conference. Each of the interviewers' avatars had a full-body presentation, but each of the 19 interviewees only had both hands as a physical presence in the virtual environment. 
In the experiments under \textit{REAL} condition, the interviewee had a face-to-face interview with a live interviewer (see \autoref{fig:interview}).

The second independent variable (II) \textsc{Type of Interview Questions} corresponded to two categories of questions: professional inquiries and personnel interview questions, since the previous study has also shown that different types of interview questions have an impact on interviewee's performance \cite{hartwell2019we}. The professional inquiries consisted of the professional knowledge the interviewee has learned in college, which involved computer networking, operating systems, programming and algorithms, linear algebra, database, and principles of computer composition. Each question had the corresponding correct answers and would examine the interviewee's memory, logical thinking ability, reaction speed, and mastery of professional knowledge. These professional questions were based on university final exams, internship interviews, job interviews, and interview questions for the graduate school review. Since there was no established standard for what kind of job the participants were applying for, the personnel questions used in the simulated interviews were general and typical job interview questions. The questions were divided into five categories: basic personal information, personality assessment, emotional control, organizing and planning skills, and creative questions. The pressure and difficulty of the questions in these five categories increased, and they were eventually divided into four sets of ten questions each, with a similar level of difficulty and no repetitions. The answers to these questions were mainly based on the interviewee's review and summary of their experience and evaluation of themselves and were open-ended.
Moreover, for each of the eight groups of interviews, we prepared different interview questions accordingly, so there were eight separate sets of interview questions in total, and each category(i.e., professional and personnel) had four sets of interview questions of the same difficulty with each set including ten interview questions. 
All the interview questions were in the additional materials.

The third independent variable (III) \textsc{Interviewer attitide} corresponded to the interviewer's attitude(mainly body language and tone of voice) and response to the interviewee's performance during the interview process. A previous study discovered that the participants exhibited more anxiety by the attitude of virtual avatars than the avatar's level of realism \cite{kwon2009study}. 
We designed two types of interviewers with positive and negative attitudes. Both interviewers would give interviewees real-time responses based on their performances. The positive interviewer would respond with positive feedback on the interviewee's answers (If the interviewee did well, the interviewer would respond ``Excellent, exactly right." If the interviewee did not perform well, the interviewer would reply ``It's okay, there is no rush, please take your time to think about it.") with positive animations ( e.g., greeting, handshaking, listening with full attention, nodding, acknowledging, see \autoref{fig:realism}). 
The negative interviewer would start the interview by emphasizing ``I will only ask all the questions once and will not repeat them, so listen carefully." During an interview, the interviewer would give negative feedback on the interviewee's answers (If the interviewee is unable to answer or answers incorrectly, the interviewer will respond ``You can't answer such a simple question?” or ``Totally wrong, it's all learned knowledge.” or ``Organize your language more clearly, time is up.") , and with negative animations (e.g., shaking head, yawning, pouting, rubbing shoulders, looking around impatiently, talking on the phone, or texting, see \autoref{fig:realism}).

The fourth independent variable (IV) \textsc {Timed or Untimed Answers} corresponded to whether to time each interviewee's answer. There is also past literature on the effects of timed and untimed questions on student performance and anxiety \cite{schwartz2015comparison,morris1969effects}. 
In the case of a task without time limitation (i.e., no timing), each interview question could be answered for any length of time; in the case of a time-sensitive task (i.e., timing), for each interview question, the interviewer would time the interviewee for 30 seconds and interrupt the interviewer's answer as soon as the time is up. 

The fifth independent variable (V) \textsc{With or Without Preparation} corresponded to whether the interviewee was given 5 minutes to prepare for that round of the interview before the job interview began. During the 5 minutes, the interviewee could review the ten interview questions for that round and could search for information or memorize the relevant materials distributed by the staff to structure their answers in advance.

\begin{table*}[h]
\caption{Orthogonal design with multi-factors and mixed levels.}\label{tab:experiment}
\centering
\begin{tabular}{cllccc}
\toprule
\multicolumn{1}{l}{\textbf{\begin{tabular}[c]{@{}l@{}}Experiment\\ group\end{tabular}}} & \textbf{\begin{tabular}[c]{@{}l@{}}Interviewer\\ attitude\end{tabular}} & \multicolumn{1}{c}{\textbf{\begin{tabular}[c]{@{}c@{}}Question \\ type\end{tabular}}} & \multicolumn{1}{c}{\textbf{Timing}} & \textbf{Preparation} & \textbf{Realism}   \\ \hline
1                                                                                       & Positive                                                                & Personal                                                                     & Yes                                 & Yes                  & Video-conferencing \\
2                                                                                       & Positive                                                                & Personal                                                                     & No                                  & No                   & Realistic VR       \\
3                                                                                       & Positive                                                                & Professional                                                                 & Yes                                 & No                   & Cartoon VR         \\
4                                                                                       & Positive                                                                & Professional                                                                 & No                                  & Yes                  & Onsite meeting     \\
5                                                                                       & Negative                                                                & Personal                                                                     & Yes                                 & No                  & Onsite meeting     \\
6                                                                                       & Negative                                                                & Personal                                                                     & No                                  & Yes                   & Cartoon VR         \\
7                                                                                       & Negative                                                                & Professional                                                                 & Yes                                 & Yes                   & Realistic VR       \\
8                                                                                       & Negative                                                                & Professional                                                                 & No                                  & No                  & Video-conferencing \\
\bottomrule
\end{tabular}
\end{table*}

%We investigated the causes of anxiety by measuring electrodermal activities and heart rate during a job interview.
\subsection{Apparatus}
\autoref{fig:interview} shows our experiment setup. The whole system had four different settings for the four levels of realism. In all conditions,  participants were asked to wear an E4 wristband on their left wrist, and a smartphone on the side would display the real-time physiological data acquisition without being seen by the participant. 
For the video-conferencing condition, interviewers conducted video conferences with the interviewee via Tencent meeting on a laptop computer with Windows 10 operating system, NVIDIA GeForce RTX 3060 GPU, and a 16.1'' monitor with a resolution of $1920 \times 1080$. 
The cartoon VR and realistic VR conditions used the same experimental equipment setup. Interviewers conducted a Tencent meeting with the interviewee on a laptop computer while wearing a Meta Oculus Quest 2, a standalone headset with an internal, Android-based operating system, graphics of $1832 \times 1920$ pixels per eye at 90 Hz, and a 6 GB of LPDDR4X RAM processor. Through a fiber-optic link cable, we connected the socket of the VR headset to the USB socket of the laptop computer, which allowed us to cast the scene rendered in the VR headset directly to the laptop computer through the SideQuest application. Then, the picture on the laptop computer would be screen shared with the interviewer through Tencent meeting so that the interviewer could give verbal feedback in real-time according to the animations of the avatar and the interviewee's performance. The interviewee could also hear the interviewer's voice reply in the Tencent meeting on the laptop computer linked to the VR headset.
For the onsite meeting condition, interviewees had a face-to-face interview with real human interviewers in the real site as \autoref{fig:scene} demonstrated.

\subsection{Application}
\begin{figure}[ht]
\centering
\includegraphics[width=0.9\linewidth]{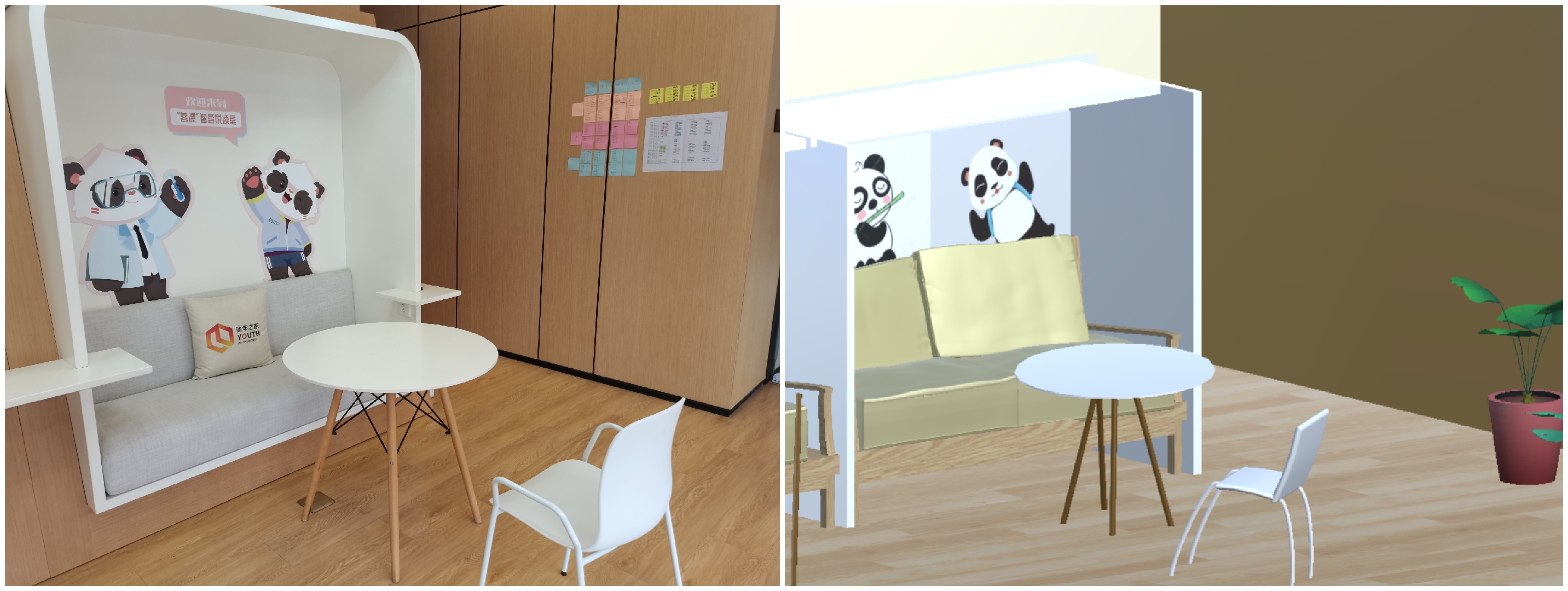}
\caption{Real interview site (L) \& virtual scenario built in Unity (R).}
\label{fig:scene}
\end{figure}

The Unity applications consisted of scenes and interviewers' avatars, and the Unity3D version we used for developing the application is 2020.3.25. In order to focus only on the realism of the avatar itself, we excluded the interference of different environments by making them the same as the physical environment. We built a virtual interview scene based on a real interview scene by using the abundant 3D models in the Unity Asset Store. \autoref{fig:scene} demonstrates real interview site and virtual interview scenario. We designed the corresponding avatars based on two real females (see \autoref{fig:realism}). In order to present
avatars with different realism (cartoon and realistic), we used two different modeling approaches. The cartoon avatar was developed by using \textit{Ready Player Me}\footnote{\url{https://readyplayer.me}}, a free web platform that supports users to automatically generate an avatar that resembles a real person by uploading a selfie. The realistic avatar was created by using \textit{Avatar SDK}\footnote{\url{https://avatarsdk.com}}, which is an advanced avatar creation toolkit using AI to create photorealistic and lifelike 3D avatars from selfie photos. In order to rig the skeleton and animate the avatars, we used the Mixamo auto rigging tool to rig and animate our characters by uploading the models to the Mixamo website and selecting the animations needed by the corresponding avatars for downloads (e.g., handshaking for the positive interviewer and pouting for the negative interviewer). Avatars of interviewers were programmed to achieve various
autonomous animations using the corresponding animator controller.
With the animator components added to our avatars, the avatars would automatically play the pre-customized animation sequence with natural transitions. Using the Unity XR Interaction Toolkit \footnote{\url{https://docs.unity3d.com/Packages/com.unity.xr.interaction.toolkit@2.2/manual/index.html}},
we built our application on the Android platform as ``apk" files, which would run on an Oculus Quest 2 headset.

%\begin{figure*}[ht]
%\centering
%\includegraphics[width=0.4\columnwidth,angle=0]{./figure/architecture.jpg}
%\caption{A.......}
%\label{fig:model} %
%\end{figure*}

\subsection{Participants}

The participants were recruited from our university campuses with voluntary consensus. They were undergraduate students facing internship, job, and graduate school review interviews in the next year or two. They were, therefore, likely to be the primary users of the VRIS.
Nineteen university students (M = 11, F = 8) participated in this experiment, aged between 19 and 21 (M =19.9, SD=0.64). Participants received \textyen100 each for their participation.
Each participant performed the interviews in all eight experiment conditions within four days (twice daily). Each participant was interviewed twice a day (i.e., once in the morning and once in the evening) with a 12-hour interval between the two sessions. Each interview is approximately 5-10 minutes long. The serial numbers of the experiment conducted each time were counterbalanced to reduce sequence effects.

\subsection{Procedure}
\begin{figure}[ht]
\centering
\includegraphics[width=0.9\linewidth]{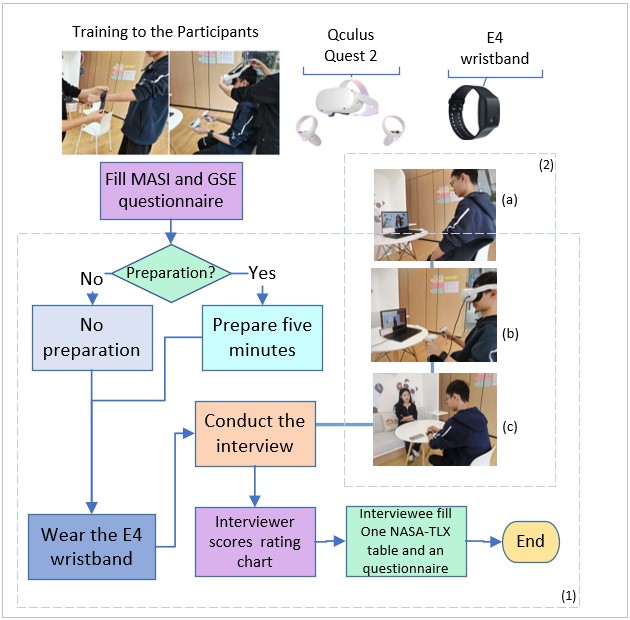}
\caption{Flow chart of the experiment. (1) Repeat two times in one day (one in the morning and one in the evening with a 12-hour interval). Each participant performed the interviews in all 8 different experiment conditions within 4 days. (2) Choose one of the experiment setup for each interview. }
\label{fig:flow_chart}
\end{figure}

The flow chart of the experimental protocol is represented in \autoref{fig:flow_chart}.
%Experiments are performed in a controlled studio environment and is also set up like a natural and quiet lounge where users are likely to use VRIS to simulate their job interviews.
Experiments were conducted in a controlled studio setting where the room temperature was set to 25℃-27℃ with indoor air conditioning. The studio was also designed to resemble a pleasant, calm lounge where users may use VRIS to mimic job interviews. We supposed that many real-life interviews would take place in a similar scenario. 
Before all interviews began, participants were asked to fill out the Measure of Anxiety in Selection Interviews (MASI) and General Self-Efficacy (GSE) scale in order to filter out exceptional cases of being too nervous about the interview (or even suffering from a related illness) or not nervous at all.
In our interview experiment, participants were told to imagine that this was a real job interview scenario and that each interview question asked by the interviewer needed to be thought through and answered carefully. 
During the interview, a staff member sitting beside the interviewee would collect physiological data from the bracelet.
At the end of each interview, the interviewees were asked to fill out two questionnaires (a NASA Task Load Index (NASA TLX) and a self-assessment anxiety questionnaire) based on their experience during the interview they just completed, and the interviewer will rate the interviewees' performance on a score sheet for the round.

% \subsection{Measure of anxiety}
\subsection{Data Analysis}
In order to comprehensively assess interviewer anxiety, overall experience, and interview performance, we combined subjective questionnaires and objective physiological signals, where the questionnaires were divided into interviewees' self-perceptions and interviewers' observations and evaluations; and we used electrodermal activity, in particular skin conductance response, as a quantitative approach to measure interviewees' anxiety.
% \subsubsection{Self-rating questionnaires}
% \paragraph{3.5.1.1 Measures Administered to Interviewees}

\textit{\textbf{Questionnaires: }}Interviewees were asked to fill in a paper version of the questionnaires to measure self-rated anxiety both before and after the interview. Before starting the experiment, each interviewee was asked to fill in personal information, the Measure of Anxiety in Selection Interviews (MASI), and the General Self-Efficacy (GSE) scale. Then after each set of experiments, interviewees were asked to complete the NASA Task Load Index (TLX) and an anxiety self-assessment questionnaire.

The Measure of Anxiety in Selection Interviews (MASI) was used to measure interviewees' self-rated interview anxiety. MASI is a concise and practical measurement tool that comprehensively assesses multiple aspects of job interview anxiety \cite{mccarthy2004measuring}. The MASI includes measures of interview anxiety across five dimensions: communication anxiety, appearance anxiety, social anxiety, performance anxiety, and behavioral anxiety. Each dimension includes six questions, for a total of 30 questions.  More than 35\% of the answers to the MASI questions scored above three on the five-point response scale (1= strongly disagree, 5= strongly agree), indicating that interviewees displayed considerable anxiety in at least some aspects of the interview process \cite{zielinska2021virtual}. Based on this criterion, i.e., ``11 questions with a score greater than 3'', the eligible samples of MASI results were screened and aggregated, with nine interviewees experiencing substantial interview anxiety, representing 47\% of the total number of interviewees.

Self-efficacy is a central concept in Bandura's social cognitive theory, where ``efficacy expectations are presumed to influence the level of performance by enhancing intensity and persistence of effort''. Further examination revealed that self-efficacy significantly influences human behavior (e.g., stress reactions, self-regulation, coping, achievement striving, career pursuits) \cite{zeng2020evaluating}. A 10-item version of the General Self-Efficacy (GSE) scale \cite{johnston1995measures} of which the total score range is 10-40 using a 5-point Likert metric with options ``1 = Not at all true, 2 = Hardly true, 3 = Moderately true, 4 = Exactly true'' was used to measure the interviewees' ability to deal with various stressors in their life and, in particular, to have control over their actions in the interview setting. The Chinese version of GSE \cite{zhang1995measuring}has been validated by Zhang and Schwarzer to have good reliability and validity. People with different levels of self-efficacy feel, think and act differently. At the level of feelings, self-efficacy is often associated with depression, anxiety, and helplessness. Generally, a score of 20 or below on the GSE scale indicates low self-efficacy. After analyzing the data, we calculated that the average score of the participants was 25.21, and 18 participants scored between 20 and 30, i.e., they met
the criteria of "high self-efficacy," while only one participant scored 17 indicating "low self-efficacy and sometimes low confidence ." 

National Aeronautics and Space Administration Task Load Index (NASA TLX) 
 \cite{hart1986nasa} is a subjective workload rating scale. It uses six dimensions to assess workload: mental demand, physical demand, temporal demand, performance, effort, and frustration. Interviewees are asked to rate each dimension on a twenty-step bipolar scale with a score from 0 to 100 (0= Very Low, 100=Very High). NASA TLX has been proven to have the highest factor validity rated the best in its ability to represent workload in the four subjective workload scales \cite{hill1992comparison}. This scale measures the interview's mental, performance, and psychological effects on the interviewer across different experimental groups.

We also designed a self-assessment questionnaire with three questions for interviewees to rate their levels of anxiety, subjective discomfort, and eye contact avoidance, as studies have found that less eye contact is strongly associated with the interviewee being more anxious, more uncomfortable and less well behaved, which equates to more nervousness \cite{howell2016relations}.
The scale of both anxiety and discomfort ranged from 0 to 100, and interviewees were given four options to measure their eye contact: no eye contact, glance, gaze, or gaze with a smile. Each choice was eventually mapped to a scale of 0-100 for analysis.  
% \paragraph{3.5.1.2 Measures Administered to Interviewers}
% \paragraph{Interviewers' rating}

Prior research has suggested that interviewers can detect interview anxiety with reasonable accuracy \cite{feiler2010interviewee}. Therefore, for interviewers to evaluate each interview session based on the interviewees' performance, a performance rating scale with three dimensions, each with a score range from 0 to 100, was applied to ascertain the participants' perceptions of their anxiety and behavior during the interview.
% The interviewers were asked to rate the interviewees in three dimensions. 
First, the interviewee's nervousness was scored through non-verbal behaviors, such as shaking hands, stiff movements, or demonstrating little eye contact. Second, the interviewee's performance was scored on their answers' accuracy, logic, and time control. Third, the interviewee's communication skill is scored by the ratio of pauses, errors, stammering, slurring, and verbal chanting throughout the interview.
The self-assessment questionnaire and performance rating scale can be found in the appendix.

\textbf{\textit{Physiology Measures: }}Physiological measures related to feelings of anxiety are mainly composed of EDA(Electrodermal Activity). We used the Empatica E4 bracelet with two electrode sensors to measure these data. It captures conductance (inverse resistance) through the skin, passing a minimal amount of current between two electrodes in contact with the skin to obtain EDA signal data. When experiencing emotional activation, increased cognitive load, or physical exertion, the brain sends signals to the skin, the pores begin to fill below the surface, and conductance increases in a measurable way \cite{greco2016advances}. One component of EDA is the phasic component, which refers to the faster-changing elements of the signal - the Skin Conductance Response (SCR)~\cite{braithwaite2013guide}.
Tonic data in EDA signals such as SCL(Skin conductance level) levels vary according to individual differences and changes in the experimental setting and thus need to record baseline for further analysis. Yet, our analysis mainly focused on ER-SCR(Event-related skin conductance response), which is one kind of phasic data
when specific events (e.g., visual stimuli or stressful events) induce corresponding SCRs where individual differences and changes in time and environment play little role. So recording the baseline is not mandatory when analyzing ER-SCR in our experiment. 
The processing for extracting the ER-SCR from the EDA data is as follows:\\
(1) Raw data collection: The data from the E4 wristband website was collected right after each interview. And then, raw EDA data with Unix timestamps were converted to the local time of the lab.\\
(2) ER-SCR extraction: we used the neurokit2 package\footnote{\url{https://pypi.org/project/neurokit2/}}, a python toolbox for neurophysiological signal processing. Through functions neurokit2 provided, we could feed raw EDA signal and got returns like the number of occurrences of Skin Conductance Response (SCR), the mean amplitude of the SCR peak occurrences, and other SCR information for future analysis \cite{makowski2021neurokit2}.

\section{Results}

All participants completed the orthogonal experiment, and therefore we opted for the mixed-effects model (also named ``multilevel model'' or ``hierarchical model'') to analyze the data from repeated measures~\cite{bates2014fitting}. The mixed-effects model includes fixed effects associated with the response variable and random effects not related to the response variable. In order to get over the effect of individual differences on the interview experience, we set it as a random effect in this model. Other variables presented in \autoref{tab:experiment} are fixed effects. Mixed-effects model can study whether one factor has a higher impact on the response variable compared to the other factors. In comparison, the post-hoc analysis can compare the effect of different levels in one factor. Therefore, we carry out the analysis of the mixed-effects model and post-hoc analysis. Data analysis was performed in $\mathit{R}$, and no significant interaction effects of any combinations of independent variables were detected. The significance level was set to $.05$. We also used corrections when performing the post-hoc analysis.

% \subsection{Interviewee's feedback}
\subsection{Feedback on Anxiety}

\autoref{tab:anxiety} presents the results of mixed-effects model for analyzing the effect of five independent variables (\textit{Interviewer attitude, Question type, Time keeping, Preparation, Realism}) on interviewees' anxiety. We measured interviewees' anxiety in three ways, which were 
\textbf{self-perceived anxiety}, \textbf{physiological anxiety} and \textbf{interviewer-rated anxiety};
\textbf{self-perceived anxiety} was from the self-rated anxiety scores in questionnaires for interviewees as a subjective indicator, \textbf{physiological anxiety} was the ER-SCR data extracted from electrodermal activities embodying the level of anxiety as an objective indicator, and \textbf{interviewer-rated anxiety} was from the rating chart for interviewers.
Results from mixed-model rANOVA indicated that
\textit{Question type}~($F_{1,133}=89.31,p=.01,\eta_p^2=.40$), \textit{Preparation}~($F_{1,133}=5.45,p=.02,\eta_p^2=.04$) and \textit{Realism}~($F_{1,133}=3.14,p=.02,\eta_p^2=.02$) all have significant influence on \textbf{self-perceived anxiety}; while \textit{Question type}~($F_{1,133}=4.30,p=.04,\eta_p^2=.03$) and \textit{Interviewer attitude}~($F_{1,133}=4.38,p=.04,\eta_p^2=.03$) both significantly affect \textit{physiological anxiety}; followed with \textbf{interviewer-rated anxiety}, from the interviewer's point, interviewees' anxiety is affected by \textit{Question type}~($F_{1,133}=22.61,p<.01,\eta_p^2=.15$), \textit{Interviewer attitude}~($F_{1,133}=6.78,p=.01,\eta_p^2=.05$), and \textit{Time keeping}~($F_{1,133}=10.04,p<.01,\eta_p^2=.07$), which is quite different to the results reported by interviewees.

\begin{table}[!ht]
\caption{Analysis of factors that affecting interviewee's anxiety using mixed-effects model(NumDF=1, DenDF=133)}
\label{tab:anxiety}
\centering
\scalebox{0.5}{
\begin{tabular}{cllllll}
\hline
                                                               & \textbf{\begin{tabular}[c]{@{}l@{}}Interviewer\\ attitude\end{tabular}}             & \textbf{\begin{tabular}[c]{@{}l@{}}Question\\ type\end{tabular}}                        & \textbf{\begin{tabular}[c]{@{}l@{}}Time\\ keeping\end{tabular}}                    & \textbf{Preparation}                                                                 & \textbf{Realism}                                                                     \\ \hline
\begin{tabular}[c]{@{}l@{}}Self-perceived\\Anxiety  \end{tabular}
                                                      & \begin{tabular}[c]{@{}l@{}}SS=55.7\\ F=.23\\ p=.63\\ $\eta_p^2$=.40\end{tabular}    & \begin{tabular}[c]{@{}l@{}}SS=22032.2\\ F=89.31\\ \textbf{p\textless{}.01**}\\ $\eta_p^2$=.00\end{tabular} & \begin{tabular}[c]{@{}l@{}}SS=1.3\\ F=.005\\ p=.94\\ $\eta_p^2$=.00\end{tabular}   & \begin{tabular}[c]{@{}l@{}}SS=1344.1\\ F=5.45\\ \textbf{p=.02*}\\ $\eta_p^2$=.04\end{tabular} & \begin{tabular}[c]{@{}l@{}}SS=2322.7\\ F=3.14\\ \textbf{p=.02*}\\ $\eta_p^2$=.02\end{tabular} \\ \hline
\begin{tabular}[c]{@{}l@{}}ER-SCR\\  (Gomboa2008)\end{tabular} & \begin{tabular}[c]{@{}l@{}}SS=.0074\\ F=4.38\\ \textbf{p=.04*}\\ $\eta_p^2$=.03\end{tabular} & \begin{tabular}[c]{@{}l@{}}SS=.0072\\ F=4.30\\ \textbf{p=.04*}\\ $\eta_p^2$=.03\end{tabular}     & \begin{tabular}[c]{@{}l@{}}SS=.0034\\ F=2.05\\ p=.15\\ $\eta_p^2$=.02\end{tabular} & \begin{tabular}[c]{@{}l@{}}SS=.0005\\ F=.32\\ p=.57\\ $\eta_p^2$=.00\end{tabular}    & \begin{tabular}[c]{@{}l@{}}SS=.0014\\ F=.28\\ p=.84\\ $\eta_p^2$=.00\end{tabular}    \\ \hline
\begin{tabular}[c]{@{}l@{}}Interviewer-rated\\ Anxiety\end{tabular} & \begin{tabular}[c]{@{}l@{}}SS=1392.1\\ F=6.78\\ \textbf{p=.01*}\\ $\eta_p^2$=.05\end{tabular} & \begin{tabular}[c]{@{}l@{}}SS=4642.1\\ F=22.61\\ \textbf{p$<$.01**}\\ $\eta_p^2$=.15\end{tabular}     & \begin{tabular}[c]{@{}l@{}}SS=2063.2\\ F=10.04\\ \textbf{p$<$.01**}\\ $\eta_p^2$=.07\end{tabular} & \begin{tabular}[c]{@{}l@{}}SS=65.8\\ F=.32\\ p=.57\\ $\eta_p^2$=.00\end{tabular}    & \begin{tabular}[c]{@{}l@{}}SS=1276.3\\ F=2.07\\ p=.11\\ $\eta_p^2$=.02\end{tabular} \\ \hline
\end{tabular}
}
\end{table}

Further post-hoc analysis is shown in \autoref{tab:anxiety_post-hoc}. Pairwise comparisons were performed using the t-test. Compared to \textit{Personal question}, \textit{Professional question} significantly leads to higher \textbf{self-perceived anxiety}~($MD=24.07, t_{133}=9.45, p<.01, \eta_p^2=.40$), \textit{Without preparation} also leads to higher \textbf{self-perceived anxiety}~($MD=-5.95, t_{133}=-2.33, \textbf{p=.02*}, \eta_p^2=.04$), compared with the \textit{Realistic VR} interview, \textit{PC} interview are linked to lower \textit{self-perceived anxiety}~($MD=-7.5, t_{133}=-2.08, p=.03, \eta_p^2=.03$ ), similarly, \textit{Real person} interview leads to significantly lower \textit{self-perceived anxiety}~($MD=-10.7, t_{133}=-2.97, p<.01, \eta_p^2=.06$) than \textit{Realistic VR}.

\begin{table*}[!bp]
\centering
\caption{Post-hoc analysis for factors affecting interviewee's anxiety using t test (MD=mean difference, df=133). }
\label{tab:anxiety_post-hoc}
\scalebox{0.5}{
\begin{tabular}{cllllllllll}
\hline
                                                                          & \textbf{\begin{tabular}[c]{@{}l@{}}Interviewer attitude:\\ Negative - Positive\end{tabular}}                        & \textbf{\begin{tabular}[c]{@{}l@{}}Question type:\\ Professional - Personal\end{tabular}}                     & \textbf{\begin{tabular}[c]{@{}l@{}}Timekeeping:\\ Yes- No\end{tabular}}       & \textbf{\begin{tabular}[c]{@{}l@{}}Preparation:\\ Yes - No\end{tabular}}    & \textbf{\begin{tabular}[c]{@{}l@{}}Realism:\\ Cartoon VR -\\ PC\end{tabular}}  & \textbf{\begin{tabular}[c]{@{}l@{}}Realism:\\ Cartoon VR -\\ Real Person\end{tabular}}  & \textbf{\begin{tabular}[c]{@{}l@{}}Realism:\\ Cartoon VR -\\ Realistic VR\end{tabular}}        & \textbf{\begin{tabular}[c]{@{}l@{}}Realism:\\ PC - \\Real Person\end{tabular}}        & \textbf{\begin{tabular}[c]{@{}l@{}}Realism:\\ PC - \\Realistic VR\end{tabular}}        & \textbf{\begin{tabular}[c]{@{}l@{}}Realism:\\ Real Person - \\Realistic VR\end{tabular}}        
                                                               \\ \hline
\begin{tabular}[c]{@{}l@{}}Self-perceived\\Anxiety  \end{tabular}
                                                      & \begin{tabular}[c]{@{}l@{}}MD=1.21\\ t=.48\\ p=.64\\ $\eta_p^2$=.00\end{tabular}    
                                                      & \begin{tabular}[c]{@{}l@{}}MD=24.07\\ t=9.45\\ \textbf{p\textless{}.01**}\\ $\eta_p^2$=.40\end{tabular} 
                                                      & \begin{tabular}[c]{@{}l@{}}MD=0.18\\ t=.07\\ p=.94\\ $\eta_p^2$=.00\end{tabular}   
                                                      & \begin{tabular}[c]{@{}l@{}}MD=-5.95\\ t=-2.33\\ \textbf{p=.02*}\\ $\eta_p^2$=.04\end{tabular} 
                                                      & \begin{tabular}[c]{@{}l@{}}MD=2.39\\ t=.66\\ p=.51\\ $\eta_p^2$=.00\end{tabular} 
                                                      & \begin{tabular}[c]{@{}l@{}}MD=5.61\\ t=1.56\\ p=.12\\ $\eta_p^2$=.02\end{tabular} 
                                                      & \begin{tabular}[c]{@{}l@{}}MD=-5.11\\ t==-1.42\\ p=.15\\ $\eta_p^2$=.01\end{tabular} 
                                                      & \begin{tabular}[c]{@{}l@{}}MD=3.21\\ t=.89\\ p=.37\\ $\eta_p^2$=.00\end{tabular} 
                                                      & \begin{tabular}[c]{@{}l@{}}MD=-7.5\\ t=-2.08\\ \textbf{p=.03*}\\ $\eta_p^2$=.03\end{tabular} 
                                                      & \begin{tabular}[c]{@{}l@{}}MD=-10.7\\ t=-2.97\\ \textbf{p\textless{}.01**}\\ $\eta_p^2$=.06\end{tabular}

                                                      \\ \hline
\begin{tabular}[c]{@{}l@{}}ER-SCR\\  (Gomboa2008)\end{tabular} 
& \begin{tabular}[c]{@{}l@{}}MD=.01\\ t=2.10\\ \textbf{p=.04*}\\ $\eta_p^2$=.03\end{tabular} 
& \begin{tabular}[c]{@{}l@{}}MD=.01\\ t=2.07\\ \textbf{p=.04*}\\ $\eta_p^2$=.03\end{tabular}     
& \begin{tabular}[c]{@{}l@{}}MD=-.01\\ t=-1.43\\ p=.15\\ $\eta_p^2$=.02\end{tabular} 
& \begin{tabular}[c]{@{}l@{}}MD=-.004\\ t=-.57\\ p=.57\\ $\eta_p^2$=.00\end{tabular}    
& \begin{tabular}[c]{@{}l@{}}MD=-.007\\ t=-.81\\ p=.41\\ $\eta_p^2$=.00\end{tabular}
& \begin{tabular}[c]{@{}l@{}}MD=.003\\ t=-.32\\ p=.75\\ $\eta_p^2$=.00\end{tabular}
& \begin{tabular}[c]{@{}l@{}}MD=-.0003\\ t=-.03\\ p=.97\\ $\eta_p^2$=.00\end{tabular}  
& \begin{tabular}[c]{@{}l@{}}MD=.004\\ t=.50\\ p=.62\\ $\eta_p^2$=.00\end{tabular}
& \begin{tabular}[c]{@{}l@{}}MD=.007\\ t=.78\\ p=.44\\ $\eta_p^2$=.00\end{tabular}
& \begin{tabular}[c]{@{}l@{}}MD=.002\\ t=.28\\ p=.78\\ $\eta_p^2$=.00\end{tabular} \\ \hline

\begin{tabular}[c]{@{}l@{}}Interviewer-rated\\ Anxiety\end{tabular} 
& \begin{tabular}[c]{@{}l@{}}MD=6.05\\ t=2.60\\ \textbf{p=.01**}\\ $\eta_p^2$=.05\end{tabular} 
& \begin{tabular}[c]{@{}l@{}}MD=11.05\\ t=4.75\\ \textbf{p\textless{}.01**}\\ $\eta_p^2$=.15\end{tabular}     
& \begin{tabular}[c]{@{}l@{}}MD=7.37\\ t=3.17\\ \textbf{p\textless{}.01**}\\ $\eta_p^2$=.07\end{tabular} 
& \begin{tabular}[c]{@{}l@{}}MD=1.32\\ t=.57\\ p=.57\\ $\eta_p^2$=.00\end{tabular}    
& \begin{tabular}[c]{@{}l@{}}MD=2.89\\ t=.88\\ p=.38\\ $\eta_p^2$=.00\end{tabular} 
& \begin{tabular}[c]{@{}l@{}}MD=-2.89\\ t=-.88\\ p=.38\\ $\eta_p^2$=.00\end{tabular}
& \begin{tabular}[c]{@{}l@{}}MD=-4.74\\ t==-1.44\\ p=.15\\ $\eta_p^2$=.02\end{tabular} 
& \begin{tabular}[c]{@{}l@{}}MD=-5.79\\ t=-1.76\\ p=.08\\ $\eta_p^2$=.02\end{tabular}
& \begin{tabular}[c]{@{}l@{}}MD=-7.63\\ t=-2.32\\ \textbf{p=.02*}\\ $\eta_p^2$=.04\end{tabular}
& \begin{tabular}[c]{@{}l@{}}MD=-7.63\\ t=-2.32\\ \textbf{p=.02*}\\ $\eta_p^2$=.04\end{tabular}\\ \hline
\end{tabular}
}
\end{table*}

When it comes to \textbf{physiological anxiety}, only \textit{Question type} and \textit{Interviewer attitude} have significant influence with a negative interviewer($MD=.01, t_{133}=2.10, p=.04, \eta_p^2=.03$) inducing more \textbf{physiological anxiety} than the positive one; yet unlike \textbf{self-assessed anxiety}, personal questions~($MD=-0.01, t_{133}=-2.07, p=.04, \eta_p^2=.03$)  tends to induce more \textbf{physiological anxiety} than professional ones.

In terms of \textbf{interviewer-rated anxiety}, same as \textbf{self-assessed anxiety}, \textit{Professional question} significantly leads to higher \textbf{interviewer-rated anxiety}~($MD=11.05, t_{133}=4.75, p<.01, \eta_p^2=.15$), also \textit{Real person}($MD=-7.63, t_{133}=2.32, p=.02, \eta_p^2=.04$) and \textit{PC}($MD=-7.63, t_{133}=2.32, p=.02, \eta_p^2=.04$) interview both induce less \textbf{interviewer-rated anxiety} than \textit{Realistic VR} which is in line with the results of \textbf{self-assessed anxiety}; but inconsistent with \textbf{self-assessed anxiety}, \textit{Interviewer attitude} and \textit{Time keeping} also play a role with a negative interviewer($MD=-6.05, t_{133}=-2.60, p=.01, \eta_p^2=.05$) and keeping time for 30s each question~($MD=7.37, t_{133}=3.17, p<.01, \eta_p^2=.07$) inducing more \textbf{interviewer-rated anxiety} than the positive one and without time keeping.

\subsection{Overall Experience}
To get a full picture of the interviewee' experience, we collected their \textbf{Cognitive load}, \textbf{Discomfort} and \textbf{Avoidance of eye contact} through NASA-TLX and subjective self-assessment questionnaires. 

\autoref{tab:interviewee_nasa} demonstrates the effect of interview factors on the cognitive workload measured through the NASA-TLX in six dimension using mixed-effects model. We can observe that \textit{Interviewer attitude} doesn't have any significant influence on the NASA-TLX criteria, while \textit{Question type} has significantly affected \textbf{Mental demand}~($F_{1,133}=37.79,p<.01,\eta_p^2=.22$), \textbf{Physical demand}~($F_{1,133}=7.72,p<.01,\eta_p^2=.05$), \textbf{Temporal demand}~($F_{1,133}=20.85,p<.01,\eta_p^2=.14$), \textbf{Performance}~($F_{1,133}=33.75,p<.01,\eta_p^2=.14$), \textbf{Effort}~($F_{1,133}=3.78,p=.05,\eta_p^2=.02$), \textbf{Frustration}~($F_{1,133}=34.87,p<.01,\eta_p^2=.21$). Among six criteria, \textit{Timekeeping} only has significant effect on \textbf{Physical demand}~($F_{1,133}=4.50,p=.03,\eta_p^2=.03$) and \textbf{Temporal demand}~($F_{1,133}=15.11,p<.01,\eta_p^2=.10$). \textit{Preparation} and \textit{Realism} have similar significant effect on \textbf{Performance}~($F_{1,133}=6.57,p=.01,\eta_p^2=.05$, and $F_{1,133}=3.34,p=.02,\eta_p^2=.02$ respectively) and \textbf{Frustration}~($F_{1,133}=7.75,p<.01,\eta_p^2=.06$, and $F_{1,133}=3.83,p=.01,\eta_p^2=.03$ respectively).

\begin{table}[!h]
\centering
\caption{Analysis of cognitive workload of the interviewee using mixed-effects model (NumDF=1, DenDF=133)}
\label{tab:interviewee_nasa}
\scalebox{0.5}{
\begin{tabular}{cllllll}
\toprule
                                                               & \multicolumn{1}{c}{\textbf{\begin{tabular}[c]{@{}c@{}}Mental\\ demand\end{tabular}}}                & \multicolumn{1}{c}{\textbf{\begin{tabular}[c]{@{}c@{}}Physical\\ demand\end{tabular}}}            & \multicolumn{1}{c}{\textbf{\begin{tabular}[c]{@{}c@{}}Temporal \\ demand\end{tabular}}}            & \multicolumn{1}{c}{\textbf{Performance}}                                                            & \multicolumn{1}{c}{\textbf{Effort}}                                                    & \multicolumn{1}{c}{\textbf{Frustration}}                                                            \\ \hline
\begin{tabular}[c]{@{}c@{}}Interviewer\\ attitude\end{tabular} & \begin{tabular}[c]{@{}l@{}}SS=112.9\\ F=.37\\ p=.54\\ $\eta^2$=.00\end{tabular}                 & \begin{tabular}[c]{@{}l@{}}SS=20.63\\ F=.11\\ p=.73\\ $\eta^2$=.00\end{tabular}               & \begin{tabular}[c]{@{}l@{}}SS=0.8\\ F=.003\\ p=.95\\ $\eta^2$=.00\end{tabular}                 & \begin{tabular}[c]{@{}l@{}}SS=265.8\\ F=.87\\ p=.35\\ $\eta^2$=.00\end{tabular}                 & \begin{tabular}[c]{@{}l@{}}SS=720.8\\ F=2.85\\ p=.09\\ $\eta^2$=.02\end{tabular}   & \begin{tabular}[c]{@{}l@{}}SS=154.0\\ F=.45\\ p=.50\\ $\eta^2$=.00\end{tabular}                 \\ \hline
\begin{tabular}[c]{@{}c@{}}Question\\ type\end{tabular}        & \begin{tabular}[c]{@{}l@{}}SS=11620.0\\ F=37.79\\ \textbf{p\textless{}.01**}\\ $\eta^2$=.22\end{tabular} & \begin{tabular}[c]{@{}l@{}}SS=1428.7\\ F=7.72\\ \textbf{p\textless{}.01**}\\ $\eta^2$=.05\end{tabular} & \begin{tabular}[c]{@{}l@{}}SS=5508.1\\ F=20.85\\ \textbf{p\textless{}.01**}\\ $\eta^2$=.14\end{tabular} & \begin{tabular}[c]{@{}l@{}}SS=10263.2\\ F=33.75\\ \textbf{p\textless{}.01**}\\ $\eta^2$=.20\end{tabular} & \begin{tabular}[c]{@{}l@{}}SS=955.01\\ F=3.78\\ \textbf{p=.05*}\\ $\eta^2$=.02\end{tabular} & \begin{tabular}[c]{@{}l@{}}SS=12007.9\\ F=34.87\\ \textbf{p\textless{}.01**}\\ $\eta^2$=.21\end{tabular} \\ \hline
Timekeeping                                                    & \begin{tabular}[c]{@{}l@{}}SS=720.8\\ F=2.34\\ p=.13\\ $\eta^2$=.02\end{tabular}                & \begin{tabular}[c]{@{}l@{}}SS=833.79\\ F=4.50\\ \textbf{p=.03*}\\ $\eta^2$=.03\end{tabular}            & \begin{tabular}[c]{@{}l@{}}SS=3992.4\\ F=15.11\\ \textbf{p\textless{}.01**}\\ $\eta^2$=.10\end{tabular} & \begin{tabular}[c]{@{}l@{}}SS=857.4\\ F=2.82\\ p=.09\\ $\eta^2$=.02\end{tabular}                & \begin{tabular}[c]{@{}l@{}}SS=265.8\\ F=1.05\\ p=.30\\ $\eta^2$=.00\end{tabular}   & \begin{tabular}[c]{@{}l@{}}SS=519.5\\ F=1.51\\ p=.22\\ $\eta^2$=.01\end{tabular}                \\ \hline
Preparation                                                    & \begin{tabular}[c]{@{}l@{}}SS=628.2\\ F=2.04\\ p=.15\\ $\eta^2$=.02\end{tabular}                & \begin{tabular}[c]{@{}l@{}}SS=119.13\\ F=.64\\ p=.43\\ $\eta^2$=.00\end{tabular}              & \begin{tabular}[c]{@{}l@{}}SS=87.0\\ F=.33\\ p=.57\\ $\eta^2$=.00\end{tabular}                 & \begin{tabular}[c]{@{}l@{}}SS=1997.4\\ F=6.57\\ \textbf{p=.01*}\\ $\eta^2$=.05\end{tabular}              & \begin{tabular}[c]{@{}l@{}}SS=351.06\\ F=1.39\\ p=.24\\ $\eta^2$=.00\end{tabular}  & \begin{tabular}[c]{@{}l@{}}SS=2669.5\\ F=7.75\\ \textbf{p\textless{}.01**}\\ $\eta^2$=.06\end{tabular}   \\ \hline
Realism                                                        & \begin{tabular}[c]{@{}l@{}}SS=1026.8\\ F=1.11\\ p=.34\\ $\eta^2$=.00\end{tabular}               & \begin{tabular}[c]{@{}l@{}}SS=261.0\\ F=.47\\ p=.70\\ $\eta^2$=.00\end{tabular}               & \begin{tabular}[c]{@{}l@{}}SS=1085.0\\ F=1.37\\ p=.26\\ $\eta^2$=.01\end{tabular}              & \begin{tabular}[c]{@{}l@{}}SS=3044.0\\ F=3.34\\ \textbf{p=.02*}\\ $\eta^2$=.02\end{tabular}              & \begin{tabular}[c]{@{}l@{}}SS=673.7\\ F=.89\\ p=.44\\ $\eta^2$=.00\end{tabular}    & \begin{tabular}[c]{@{}l@{}}SS=3952.5\\ F=3.83\\ \textbf{p=.01*}\\ $\eta^2$=.03\end{tabular} \\
\bottomrule
\end{tabular}
}
\end{table}

The post-hoc analysis was also performed for the NASA-TLX criteria, given in~\autoref{tab:nasa-tlx_post-hoc}. \textit{Professional question} tends to increase more \textbf{Mental demand} ($MD= 17.49, t_{133}=6.15, p<.01, \eta_p^2=.22$), \textbf{Physical demand}~($MD= 6.13, t_{133}=7.32, p<.01, \eta_p^2=.29$), \textbf{Temporal demand}~($MD= 12.04, t_{133}=4.57, p<.01, \eta_p^2=.14$), \textbf{Effort}~($MD= 5.01, t_{133}=1.94, p=.05, \eta_p^2=.03$), and \textbf{Frustration}~($MD= 17.78, t_{133}=5.91, p<.01, \eta_p^2=.21$), but reduce \textbf{Performance}~($MD= 12.04, t_{133}=4.57, p<.01, \eta_p^2=.14$). With \textit{Timekeeping}, interviewee experienced more \textbf{Physical Demand}~($MD= 4.68, t_{133}=.18, p=.03, \eta_p^2=.14$) and more \textbf{Temporal demand}~($MD= 10.25, t_{133}=3.89, p<.01, \eta_p^2=.10$). Also, \textit{Preparation} can significantly improve the \textbf{Performance}~($MD= 7.75, t_{133}=2.56, p<.01, \eta_p^2=.05$) and reduce \textbf{Frustration}~($MD= -8.38, t_{133}=-2.78, p<.01, \eta_p^2=.05$). Compared to the interview via \textit{Real person}, \textit{Cartoon VR} interview has an adverse effect on the \textit{Performance}~($MD= -11.95, t_{133}=-2.99, p<.01, \eta_p^2=.06$), \textit{Realistic VR} will increase the \textbf{Frustration}~($MD= 14.13, t_{133}=-3.32, p<.01, \eta_p^2=.08$).

\begin{table}[!ht]
\centering
\caption{Post-hoc analysis for factors affecting interviewee's cognitive workload using t test (MD=mean difference, df=133). }
\label{tab:nasa-tlx_post-hoc}
\scalebox{0.45}{
\begin{tabular}{cllllll}
\toprule
\multicolumn{1}{l}{}                                                                  & \multicolumn{1}{c}{\textbf{\begin{tabular}[c]{@{}c@{}}Mental\\ demand\end{tabular}}}           & \multicolumn{1}{c}{\textbf{\begin{tabular}[c]{@{}c@{}}Physical\\ demand\end{tabular}}}        & \multicolumn{1}{c}{\textbf{\begin{tabular}[c]{@{}c@{}}Temporal\\ demand\end{tabular}}}         & \multicolumn{1}{c}{\textbf{Performance}}                                                         & \multicolumn{1}{c}{\textbf{Effort}}                                                 & \textbf{Frustration}                                                                             \\ \hline
\begin{tabular}[c]{@{}c@{}}Interviewer\\ attitude:\\ Negative - Positive\end{tabular} & \begin{tabular}[c]{@{}l@{}}MD=-1.72\\ t=-.61\\ p=.55\\ $\eta_p^2$=.00\end{tabular}             & \begin{tabular}[c]{@{}l@{}}MD=-0.74\\ t=1.78\\ p=.74\\ $\eta_p^2$=.02\end{tabular}            & \begin{tabular}[c]{@{}l@{}}MD=-0.14\\ t=-.05\\ p=.96\\ $\eta_p^2$=0\end{tabular}               & \begin{tabular}[c]{@{}l@{}}MD=-2.64\\ t=-.94\\ p=.35\\ $\eta_p^2$=.00\end{tabular}               & \begin{tabular}[c]{@{}l@{}}MD=-4.36\\ t=-1.69\\ p=.09\\ $\eta_p^2$=.02\end{tabular} & \begin{tabular}[c]{@{}l@{}}MD=2.01\\ t=-.67\\ p=.50\\ $\eta_p^2$=.00\end{tabular}                \\ \hline
\begin{tabular}[c]{@{}c@{}}Question\\ type:\\ Professional - Personal\end{tabular}    & \begin{tabular}[c]{@{}l@{}}MD=17.49\\ t=6.15\\ \textbf{p\textless{}.01**}\\ $\eta_p^2$=.22\end{tabular} & \begin{tabular}[c]{@{}l@{}}MD=6.13\\ t=7.32\\ \textbf{p\textless{}.01**}\\ $\eta_p^2$=.29\end{tabular} & \begin{tabular}[c]{@{}l@{}}MD=12.04\\ t=4.57\\ \textbf{p\textless{}.01**}\\ $\eta_p^2$=.14\end{tabular} & \begin{tabular}[c]{@{}l@{}}MD=-16.43\\ t=-5.81\\ \textbf{p\textless{}.01**}\\ $\eta_p^2$=.20\end{tabular} & \begin{tabular}[c]{@{}l@{}}MD=5.01\\ t=1.94\\ \textbf{p=.05*}\\ $\eta_p^2$=.03\end{tabular}  & \begin{tabular}[c]{@{}l@{}}MD=17.78\\ t=5.91\\ \textbf{p\textless{}.01**}\\ $\eta_p^2$=.21\end{tabular}   \\ \hline
\begin{tabular}[c]{@{}c@{}}Timekeeping:\\ Yes- No\end{tabular}                        & \begin{tabular}[c]{@{}l@{}}MD=4.35\\ t=1.53\\ p=.13\\ $\eta_p^2$=.02\end{tabular}              & \begin{tabular}[c]{@{}l@{}}MD=4.68\\ t=.18\\ \textbf{p=.03*}\\ $\eta_p^2$=.14\end{tabular}             & \begin{tabular}[c]{@{}l@{}}MD=10.25\\ t=3.89\\ \textbf{p\textless{}.01**}\\ $\eta_p^2$=.10\end{tabular} & \begin{tabular}[c]{@{}l@{}}MD=-4.75\\ t=-1.68\\ p=.10\\ $\eta_p^2$=.02\end{tabular}              & \begin{tabular}[c]{@{}l@{}}MD=-2.64\\ t=-1.03\\ p=.31\\ $\eta_p^2$=.00\end{tabular} & \begin{tabular}[c]{@{}l@{}}MD=3.70\\ t=1.23\\ p=.22\\ $\eta_p^2$=.01\end{tabular}                \\ \hline
\begin{tabular}[c]{@{}c@{}}Preparation:\\ Yes - No\end{tabular}                       & \begin{tabular}[c]{@{}l@{}}MD=-4.06\\ t=-1.43\\ p=.15\\ $\eta_p^2$=.02\end{tabular}            & \begin{tabular}[c]{@{}l@{}}M=1.76\\ t=-1.95\\ p=.42\\ $\eta_p^2$=.03\end{tabular}             & \begin{tabular}[c]{@{}l@{}}MD=-1.51\\ t=-.57\\ p\textless{}.57\\ $\eta_p^2$=0\end{tabular}     & \begin{tabular}[c]{@{}l@{}}MD=7.25\\ t=2.56\\ \textbf{p=.01**}\\ $\eta_p^2$=.05\end{tabular}              & \begin{tabular}[c]{@{}l@{}}MD=3.04\\ t=1.18\\ p=.24\\ $\eta_p^2$=.01\end{tabular}   & \begin{tabular}[c]{@{}l@{}}MD=-8.38\\ t=-2.78\\ \textbf{p\textless{}.01**}\\ $\eta_p^2$=.05\end{tabular}  \\ \hline
\begin{tabular}[c]{@{}c@{}}Realism:\\ Cartoon VR - PC\end{tabular}                    & \begin{tabular}[c]{@{}l@{}}MD=1.45\\ t=.36\\ p=.72\\ $\eta_p^2$=.00\end{tabular}               & \begin{tabular}[c]{@{}l@{}}MD=-0.89\\ t=1.78\\ p=.77\\ $\eta_p^2$=.00\end{tabular}            & \begin{tabular}[c]{@{}l@{}}MD=1.61\\ t=.43\\ p=.67\\ $\eta_p^2$=0\end{tabular}                 & \begin{tabular}[c]{@{}l@{}}MD=-7.37\\ t=-1.84\\ p=.07\\ $\eta_p^2$=.02\end{tabular}              & \begin{tabular}[c]{@{}l@{}}MD=-4.87\\ t=-1.34\\ p=.18\\ $\eta_p^2$=.01\end{tabular} & \begin{tabular}[c]{@{}l@{}}MD=2.05\\ t=.48\\ p=.63\\ $\eta_p^2$=.00\end{tabular}                 \\ \hline
\begin{tabular}[c]{@{}c@{}}Realism:\\ Cartoon VR - Real Person\end{tabular}           & \begin{tabular}[c]{@{}l@{}}MD=3.42\\ t=.85\\ p=.40\\ $\eta_p^2$=.00\end{tabular}               & \begin{tabular}[c]{@{}l@{}}MD=2.66\\ t=5.06\\ p=.40\\ $\eta_p^2$=.16\end{tabular}             & \begin{tabular}[c]{@{}l@{}}MD=0.61\\ t=.16\\ p=.87\\ $\eta_p^2$=0\end{tabular}                 & \begin{tabular}[c]{@{}l@{}}MD=-11.95\\ t=-2.99\\ \textbf{p\textless{}.01**}\\ $\eta_p^2$=.06\end{tabular} & \begin{tabular}[c]{@{}l@{}}MD=-3.84\\ t=-1.05\\ p=.29\\ $\eta_p^2$=.00\end{tabular} & \begin{tabular}[c]{@{}l@{}}MD=6.66\\ t=1.56\\ p=.12\\ $\eta_p^2$=.02\end{tabular}                \\ \hline
\begin{tabular}[c]{@{}c@{}}Realism:\\ Cartoon VR - Realistic VR\end{tabular}          & \begin{tabular}[c]{@{}l@{}}MD=-3.68\\ t=-.91\\ p=.36\\ $\eta_p^2$=.00\end{tabular}            & \begin{tabular}[c]{@{}l@{}}MD=0.82\\ t=-0.74\\ p=.79\\ $\eta_p^2$=.00\end{tabular}            & \begin{tabular}[c]{@{}l@{}}MD=-5.29\\ t=-1.42\\ p=.16\\ $\eta_p^2$=.01\end{tabular}            & \begin{tabular}[c]{@{}l@{}}MD=-3.29\\ t=-.82\\ p=.41\\ $\eta_p^2$=.00\end{tabular}               & \begin{tabular}[c]{@{}l@{}}MD=-5.37\\ t=-1.47\\ p=.14\\ $\eta_p^2$=.02\end{tabular} & \begin{tabular}[c]{@{}l@{}}MD=-7.47\\ t=-1.76\\ p=.08\\ $\eta_p^2$=.02\end{tabular}              \\ \hline
\begin{tabular}[c]{@{}c@{}}Realism:\\ PC - Real Person\end{tabular}                   & \begin{tabular}[c]{@{}l@{}}MD=1.97\\ t=.49\\ p=.62\\ $\eta_p^2$=.00\end{tabular}               & \begin{tabular}[c]{@{}l@{}}MD=3.55\\ t=1.49\\ p=.25\\ $\eta_p^2$=.02\end{tabular}             & \begin{tabular}[c]{@{}l@{}}MD=-1.00\\ t=-0.27\\ p=.79\\ $\eta_p^2$=.00\end{tabular}            & \begin{tabular}[c]{@{}l@{}}MD=-4.58\\ t=-1.14\\ p=.25\\ $\eta_p^2$=.00\end{tabular}              & \begin{tabular}[c]{@{}l@{}}MD=1.03\\ t=.28\\ p=.78\\ $\eta_p^2$=.02\end{tabular}    & \begin{tabular}[c]{@{}l@{}}MD=4.61\\ t=1.08\\ p=.28\\ $\eta_p^2$=.00\end{tabular}                \\ \hline
\begin{tabular}[c]{@{}c@{}}Realism:\\ PC - Realistic VR\end{tabular}                  & \begin{tabular}[c]{@{}l@{}}MD=-5.13\\ t=-1.27\\ p=.20\\ $\eta_p^2$=.01\end{tabular}            & \begin{tabular}[c]{@{}l@{}}MD=1.71\\ t=-2.52\\ p=.58\\ $\eta_p^2$=.05\end{tabular}            & \begin{tabular}[c]{@{}l@{}}MD=-6.89\\ t=-1.85\\ p=.07\\ $\eta_p^2$=.03\end{tabular}            & \begin{tabular}[c]{@{}l@{}}MD=4.08\\ t=1.02\\ p=.31\\ $\eta_p^2$=.00\end{tabular}                & \begin{tabular}[c]{@{}l@{}}MD=-0.50\\ t=-.14\\ p=.89\\ $\eta_p^2$=.02\end{tabular}  & \begin{tabular}[c]{@{}l@{}}MD=-9.53\\ t=-2.24\\ \textbf{p=.03*}\\ $\eta_p^2$=.04\end{tabular}             \\ \hline
\begin{tabular}[c]{@{}c@{}}Realism:\\ Real Person - Realistic VR\end{tabular}         & \begin{tabular}[c]{@{}l@{}}MD=-7.10\\ t=-1.76\\ p=.08\\ $\eta_p^2$=.02\end{tabular}            & \begin{tabular}[c]{@{}l@{}}MD=-1.84\\ t=-4.01\\ p=.56\\ $\eta_p^2$=.11\end{tabular}           & \begin{tabular}[c]{@{}l@{}}MD=-5.89\\ t=-1.58\\ p=.12\\ $\eta_p^2$=.21\end{tabular}            & \begin{tabular}[c]{@{}l@{}}MD=8.66\\ t=2.16\\ \textbf{p=.03*}\\ $\eta_p^2$=.03\end{tabular}               & \begin{tabular}[c]{@{}l@{}}MD=-1.53\\ t=-.42\\ p=.68\\ $\eta_p^2$=.00\end{tabular}  & \begin{tabular}[c]{@{}l@{}}MD=-14.13\\ t=-3.32\\ \textbf{p\textless{}.01**}\\ $\eta_p^2$=.08\end{tabular} \\
\bottomrule
\end{tabular}
}
\end{table}

\autoref{tab:discomfort_and_eye contact} presents the self-perceived discomfort and the level of eye contact avoidance.

\textbf{Discomfort} is significantly influenced by \textit{Question type}($F_{1,133}=53.62,p<.01,\eta_p^2=.29$), \textit{Preparation}($F_{1,133}=3.81,p=.05,\eta_p^2=.03$), and \textit{Realism}($F_{1,133}=6.51,p<.01,\eta_p^2=.05$), while \textbf{Avoidance of eye contact} is greatly affected by \textit{Interviewer attitude}($F_{1,133}=4.03,p=.04,\eta_p^2=.03$), \textit{Question type}($F_{1,133}=8.34,p<.01,\eta_p^2=.06$), \textit{Preparation}($F_{1,133}=6.91,p<.01,\eta_p^2=.05$) and \textit{Realism}($F_{1,133}=3.69,p=.01,\eta_p^2=.03$).

\begin{table}[!h]
\centering
\caption{Analysis of interviewee's discomfort and avoidance of eye contact using mixed-effects model(NumDF=1, DenDF=133)}
\label{tab:discomfort_and_eye contact}
\scalebox{0.7}{
\begin{tabular}{cllllll}
\hline
                                                               & \textbf{\begin{tabular}[c]{@{}l@{}}Interviewer\\ attitude\end{tabular}}             & \textbf{\begin{tabular}[c]{@{}l@{}}Question\\ type\end{tabular}}                        & \textbf{\begin{tabular}[c]{@{}l@{}}Time\\ keeping\end{tabular}}                    & \textbf{Preparation}                                                                 & \textbf{Realism}                                                                     \\ \hline
\begin{tabular}[c]{@{}l@{}}Discomfort  \end{tabular}
                                                      & \begin{tabular}[c]{@{}l@{}}SS=920.2\\ F=3.17\\ p=.07\\ $\eta^2$=.02\end{tabular}   
                                                      & \begin{tabular}[c]{@{}l@{}}SS=15562.1\\ F=53.62\\ \textbf{p\textless{}.01**}\\ $\eta^2$=.29\end{tabular} & \begin{tabular}[c]{@{}l@{}}SS=9.5\\ F=.03\\ p=.85\\ $\eta^2$=.00\end{tabular}   
                                                       & \begin{tabular}[c]{@{}l@{}}SS=1105.9\\ F=3.81\\ \textbf{p=.05*}\\ $\eta^2$=.03\end{tabular} 
                                                       & \begin{tabular}[c]{@{}l@{}}SS=5673.5\\ F=6.51\\ \textbf{p\textless{}.01**}\\ $\eta^2$=.05\end{tabular}    \\ \hline
\begin{tabular}[c]{@{}l@{}}Avoidance of \\eye contact\end{tabular} & \begin{tabular}[c]{@{}l@{}}SS=1422.5\\ F=4.03\\ \textbf{p=.04*}\\ $\eta^2$=.03\end{tabular} 
&  \begin{tabular}[c]{@{}l@{}}SS=2944.5\\ F=8.34\\ \textbf{p\textless{}.01**}\\ $\eta^2$=.06\end{tabular} 
& \begin{tabular}[c]{@{}l@{}}SS=469\\ F=1.32\\ p=.25\\ $\eta^2$=0.00\end{tabular}
& \begin{tabular}[c]{@{}l@{}}SS=2440.0\\ F=6.91\\ \textbf{p\textless{}.01**}\\ $\eta^2$=.05\end{tabular}   
 & \begin{tabular}[c]{@{}l@{}}SS=3908.7\\ F=3.69\\ \textbf{p=.01*}\\ $\eta^2$=.03\end{tabular}        \\ \hline
\end{tabular}
}
\end{table}

\begin{table*}[!bp]
\caption{Post-hoc analysis for factors affecting interviewee's discomfort and eye contact using t test (MD=mean difference, df=133). }
\label{tab:disconfort_and_eye contact_post-hoc}
\centering
\scalebox{0.6}{
\begin{tabular}{cllllllllll}
\hline
                                                                          & \textbf{\begin{tabular}[c]{@{}l@{}}Interviewer attitude:\\ Negative - Positive\end{tabular}}                        & \textbf{\begin{tabular}[c]{@{}l@{}}Question type:\\ Professional - Personal\end{tabular}}                     & \textbf{\begin{tabular}[c]{@{}l@{}}Timekeeping:\\ Yes- No\end{tabular}}       & \textbf{\begin{tabular}[c]{@{}l@{}}Preparation:\\ Yes - No\end{tabular}}    & \textbf{\begin{tabular}[c]{@{}l@{}}Realism:\\ Cartoon VR - \\PC\end{tabular}}  & \textbf{\begin{tabular}[c]{@{}l@{}}Realism:\\ Cartoon VR - \\Real Person\end{tabular}}  & \textbf{\begin{tabular}[c]{@{}l@{}}Realism:\\ Cartoon VR - \\Realistic VR\end{tabular}}        & \textbf{\begin{tabular}[c]{@{}l@{}}Realism:\\ PC - Real Person\end{tabular}}        & \textbf{\begin{tabular}[c]{@{}l@{}}Realism:\\ PC - \\Realistic VR\end{tabular}}        & \textbf{\begin{tabular}[c]{@{}l@{}}Realism:\\ Real Person - \\Realistic VR\end{tabular}}        
                                                               \\ \hline
\begin{tabular}[c]{@{}l@{}}Discomfort  \end{tabular}
                                                      & \begin{tabular}[c]{@{}l@{}}MD=4.92\\ t=1.78\\ p=.07\\ $\eta_p^2$=.02\end{tabular}    
                                                      & \begin{tabular}[c]{@{}l@{}}MD=20.24\\ t=7.32\\ \textbf{p\textless{}.01**}\\ $\eta_p^2$=.29\end{tabular} 
                                                      & \begin{tabular}[c]{@{}l@{}}MD=-0.50\\ t=.18\\ p=.85\\ $\eta_p^2$=.00\end{tabular}   
                                                      & \begin{tabular}[c]{@{}l@{}}MD=-5.40\\ t=-1.95\\ \textbf{p=.05*}\\ $\eta_p^2$=.03\end{tabular} 
                                                      & \begin{tabular}[c]{@{}l@{}}MD=6.94\\ t=1.78\\ p=.078\\ $\eta_p^2$=.02\end{tabular} 
                                                      & \begin{tabular}[c]{@{}l@{}}MD=12.79\\ t=3.27\\ \textbf{p\textless{}.01**}\\ $\eta_p^2$=.07\end{tabular} 
                                                      & \begin{tabular}[c]{@{}l@{}}MD=-2.89\\ t==-0.74\\ p=.46\\ $\eta_p^2$=.06\end{tabular} 
                                                      & \begin{tabular}[c]{@{}l@{}}MD=5.84\\ t=1.49\\ p=.13\\ $\eta_p^2$=.02\end{tabular} 
                                                      & \begin{tabular}[c]{@{}l@{}}MD=-9.84\\ t=-2.52\\ \textbf{p=.01*}\\ $\eta_p^2$=.05\end{tabular} 
                                                      & \begin{tabular}[c]{@{}l@{}}MD=-15.68\\ t=-4.01\\ \textbf{p\textless{}.01**}\\ $\eta_p^2$=.11\end{tabular}

                                                      \\ \hline
\begin{tabular}[c]{@{}l@{}}Avoidance of\\ eye contact\end{tabular} 
& \begin{tabular}[c]{@{}l@{}}MD=6.12\\ t=2.01\\ \textbf{p=.04*}\\ $\eta_p^2$=.03\end{tabular} 
& \begin{tabular}[c]{@{}l@{}}MD=8.80\\ t=2.89\\ \textbf{p\textless{}.01**}\\ $\eta_p^2$=.06\end{tabular}     
& \begin{tabular}[c]{@{}l@{}}MD=3.51\\ t=1.15\\ p=.25\\ $\eta_p^2$=.00\end{tabular} 
& \begin{tabular}[c]{@{}l@{}}MD=-8.01\\ t=-2.63\\ \textbf{p\textless{}.01**}\\ $\eta_p^2$=.05\end{tabular}    
& \begin{tabular}[c]{@{}l@{}}MD=2.55\\ t=.59\\ p=.55\\ $\eta_p^2$=.00\end{tabular}
& \begin{tabular}[c]{@{}l@{}}MD=4.37\\ t=1.01\\ p=.31\\ $\eta_p^2$=.00\end{tabular}
& \begin{tabular}[c]{@{}l@{}}MD=-8.84\\ t=-2.05\\ \textbf{p=.04*}\\ $\eta_p^2$=.03\end{tabular}  
& \begin{tabular}[c]{@{}l@{}}MD=1.82\\ t=.42\\ p=.67\\ $\eta_p^2$=.00\end{tabular}
& \begin{tabular}[c]{@{}l@{}}MD=-11.40\\ t=-2.64\\ \textbf{p\textless{}.01**}\\ $\eta_p^2$=.05\end{tabular}
& \begin{tabular}[c]{@{}l@{}}MD=-13.21\\ t=-3.06\\ \textbf{p\textless{}.01**}\\ $\eta_p^2$=.07\end{tabular} \\ \hline
\end{tabular}
}
\end{table*}

Further post-hoc analysis of \textbf{Discomfort} and \textbf{Avoidance of eye contact} in \autoref{tab:disconfort_and_eye contact_post-hoc} demonstrates that more \textbf{Discomfort} discomfort tends to arise under conditions with \textit{Professional questions} than \textit{Personal questions}($MD= 20.24, t_{133}=7.32, p<.01, \eta_p^2=.29$), \textit{Without preparation} than \textit{With preparation}($MD=-5.40, t_{133}=-1.95, p=.05*, \eta_p^2=.03$), \textit{Cartoon VR} than \textit{Real person}($MD= 12.79, t_{133}=3.27, p<.01, \eta_p^2=.07$), \textit{Realistic VR} than \textit{PC}($MD= -9.84, t_{133}=-2.52, p=.01, \eta_p^2=.05$), and \textit{Realistic VR} than \textit{Real person}($MD= -15.68, t_{133}=4.01, p<.01, \eta_p^2=.11$). Also, \textit{Avoidance of eye contact} which means less eye contact with the interviewer is associated with a \textit{Negative attitude}($MD= 6.12, t_{133}=2.01, p=.04, \eta_p^2=.03$) interviewer, \textit{Personal Question}($MD= 8.80, t_{133}=2.89, p<.01, \eta_p^2=.06$) and \textit{No preparation}($MD= -8.01, t_{133}=-2.63, p<.01, \eta_p^2=.05$); surprisingly, \textit{Realistic VR} can greatly reduce eye contact than any other conditions including \textit{Cartoon VR}($MD= -8.84, t_{133}=-2.05, p=.04, \eta_p^2=.03$), \textit{PC}($MD= -11.40, t_{133}=-2.64, p<.01, \eta_p^2=.05$) and \textless{Real Person}($MD= -13.21, t_{133}=-3.06, p<.01, \eta_p^2=.07$).

\subsection{Interview Performance}
The interviewers' feedback on the interviewees' performances and ability includes \textit{Communication skill}(e.g., the ratio of pauses, errors, stammering, slurring and verbal chanting) and \textit{Overall performance}(e.g., accuracy, logic and time control); a better communication skill means more fluent, accurate and constructive oral presentation to interview questions while a better performance indicates more correct, logical and adequate answers within time limits. The mixed-effect rANOVA results are given in~\autoref{tab:overall_performance}.

From the interviewers' point, interviewers' \textit{Overall performance} can be significantly influenced by \textit{Interviewer attitude}~($F_{1,133}=3.98,p=.04,\eta_p^2=.03$), \textit{Time keeping}~($F_{1,133}=5.83,p=.02,\eta_p^2=.04$) and \textit{Preparation}~($F_{1,133}=4.91,p=.02,\eta_p^2=.04$); and their \textbf{Communication skill} tends to be affected by the \textit{Question type}~($F_{1,133}=6.41,p=.01,\eta_p^2=.05$) and \textit{Realism}~($F_{1,133}=3.05,p=.03,\eta_p^2=.02$).

\begin{table}[!h]
\centering
\caption{Analysis of interviewers' evaluation on interviewees' performance using mixed-effects model(NumDF=1, DenDF=133)}
\label{tab:overall_performance}
\scalebox{0.6}{
\begin{tabular}{cllllll}
\hline
                                                               & \textbf{\begin{tabular}[c]{@{}l@{}}Interviewer\\ attitude\end{tabular}}             & \textbf{\begin{tabular}[c]{@{}l@{}}Question\\ type\end{tabular}}                        & \textbf{\begin{tabular}[c]{@{}l@{}}Time\\ keeping\end{tabular}}                    & \textbf{Preparation}                                                                 & \textbf{Realism}                                                                     \\ \hline
\begin{tabular}[c]{@{}l@{}}Overall\\ performance  \end{tabular}
                                                      & \begin{tabular}[c]{@{}l@{}}SS=862.13\\ F=3.98\\ \textbf{p=.04*}\\ $\eta^2$=.03\end{tabular}  
                                                     & \begin{tabular}[c]{@{}l@{}}SS=.03\\ F=.0001\\ p=.99\\ $\eta^2$=.00\end{tabular} 
                                                      & \begin{tabular}[c]{@{}l@{}}SS=1262.13\\ F=5.83\\ \textbf{p=.02*}\\ $\eta^2$=.04\end{tabular} 
                                                        & \begin{tabular}[c]{@{}l@{}}SS=1063.18\\ F=4.91\\ \textbf{p=.02*}\\ $\eta^2$=.04\end{tabular} 
                                                      & \begin{tabular}[c]{@{}l@{}}SS=1203.24\\ F=1.85\\ p=.14\\ $\eta^2$=.01\end{tabular}     \\ \hline
                                                      
\begin{tabular}[c]{@{}l@{}}Communication\\ skill  \end{tabular}
                                                      & \begin{tabular}[c]{@{}l@{}}SS=716.5\\ F=3.44\\ p=.07\\ $\eta^2$=.03\end{tabular}  
                                                     & \begin{tabular}[c]{@{}l@{}}SS=1332.2\\ F=6.41\\ \textbf{p=.01*}\\ $\eta^2$=.00\end{tabular} 
                                                      & \begin{tabular}[c]{@{}l@{}}SS=111.18\\ F=.53\\ p=.46\\ $\eta^2$=.00\end{tabular} 
                                                        & \begin{tabular}[c]{@{}l@{}}SS=716.45\\ F=3.45\\ p=.06\\ $\eta^2$=.03\end{tabular} 
                                                      & \begin{tabular}[c]{@{}l@{}}SS=1902.0\\ F=3.05\\ \textbf{p=.03*}\\ $\eta^2$=.02\end{tabular}     \\ \hline
\end{tabular}
}
\end{table}

The post-hoc analysis was carried out to study the effect of different levels of significant factors from the interviewer's opinion, results are given in~\autoref{tab:performance_post-hoc}. According to the results, interviewers' \textbf{Communication skill} tends to be better with \textit{Personal question} than \textit{Professional question}($MD= -5.92, t_{133}=-2.53, p<.01, \eta_p^2=.05$), but it can be worse under condition \textit{Real Person} than \textit{Cartoon VR}($MD= 6.58, t_{133}=1.99, p<.05, \eta_p^2=.03$) and \textit{PC}($MD= 9.74, t_{133}=2.94, p<.01, \eta_p^2=.06$).
Meanwhile, interviewees' \textbf{Overall performance} is worsen by \textit{Negative interviewer}, \textit{Timekeeping}, \textit{Without preparation}, \textit{Cartoon VR}
than \textit{Positive interviewer}($MD= -4.76, t_{133}=-2.00, p=.05, \eta_p^2=.03$), \textit{No timekeeping}($MD= -5.76, t_{133}=-2.42, p=.02, \eta_p^2=.04$), \textit{With preparation}($MD= 5.29, t_{133}=2.42, p=.02, \eta_p^2=.04$), \textit{Real Person}($MD= -7.37, t_{133}=-2.18, p=.03, \eta_p^2=.03$).

\begin{table*}[!tp]
\centering
\caption{Post-hoc analysis for factors affecting interviewees' performances using t test (MD=mean difference, df=133). }
\label{tab:performance_post-hoc}
\scalebox{0.6}{
\begin{tabular}{cllllllllll}
\hline
                                                                          & \textbf{\begin{tabular}[c]{@{}l@{}}Interviewer attitude:\\ Negative - Positive\end{tabular}}                        & \textbf{\begin{tabular}[c]{@{}l@{}}Question type:\\ Professional - Personal\end{tabular}}                     & \textbf{\begin{tabular}[c]{@{}l@{}}Timekeeping:\\ Yes- No\end{tabular}}       & \textbf{\begin{tabular}[c]{@{}l@{}}Preparation:\\ Yes - No\end{tabular}}    & \textbf{\begin{tabular}[c]{@{}l@{}}Realism:\\ Cartoon VR - \\PC\end{tabular}}  & \textbf{\begin{tabular}[c]{@{}l@{}}Realism:\\ Cartoon VR - \\Real Person\end{tabular}}  & \textbf{\begin{tabular}[c]{@{}l@{}}Realism:\\ Cartoon VR-\\ Realistic VR\end{tabular}}        & \textbf{\begin{tabular}[c]{@{}l@{}}Realism:\\ PC - \\Real Person\end{tabular}}        & \textbf{\begin{tabular}[c]{@{}l@{}}Realism:\\ PC - \\Realistic VR\end{tabular}}        & \textbf{\begin{tabular}[c]{@{}l@{}}Realism:\\ Real Person - \\Realistic VR\end{tabular}}        
                                                               \\ \hline
\begin{tabular}[c]{@{}l@{}}Communication\\ skill  \end{tabular}
                                                      & \begin{tabular}[c]{@{}l@{}}MD=4.34\\ t=1.86\\ p=.07\\ $\eta_p^2$=.03\end{tabular}    
                                                      & \begin{tabular}[c]{@{}l@{}}MD=-5.92\\ t=-2.53\\ \textbf{p\textless{}.01**}\\ $\eta_p^2$=.05\end{tabular} 
                                                      & \begin{tabular}[c]{@{}l@{}}MD=1.71\\ t=.73\\ p=.47\\ $\eta_p^2$=.00\end{tabular}   
                                                      & \begin{tabular}[c]{@{}l@{}}MD=-4.34\\ t=-1.86\\ p=.07\\ $\eta_p^2$=.03\end{tabular} 
                                                      & \begin{tabular}[c]{@{}l@{}}MD=-3.16\\ t=.95\\ p=.34\\ $\eta_p^2$=.00\end{tabular} 
                                                      & \begin{tabular}[c]{@{}l@{}}MD=6.58\\ t=1.99\\ \textbf{p\textless{}.05*}\\ $\eta_p^2$=.03\end{tabular} 
                                                      & \begin{tabular}[c]{@{}l@{}}MD=-2.11\\ t==-.64\\ p=.53\\ $\eta_p^2$=.00\end{tabular} 
                                                      & \begin{tabular}[c]{@{}l@{}}MD=9.74\\ t=2.94\\ \textbf{p\textless{}.01**}\\ $\eta_p^2$=.06\end{tabular} 
                                                      & \begin{tabular}[c]{@{}l@{}}MD=5.26\\ t=1.59\\ p=.11\\ $\eta_p^2$=.02\end{tabular} 
                                                      & \begin{tabular}[c]{@{}l@{}}MD=-4.47\\ t=-1.35\\ p=.18\\ $\eta_p^2$=.01\end{tabular}

                                                      \\ \hline
\begin{tabular}[c]{@{}l@{}}Overall\\ performance\end{tabular} 
& \begin{tabular}[c]{@{}l@{}}MD=-4.76\\ t=-2.00\\ \textbf{p=.05*}\\ $\eta_p^2$=.03\end{tabular} 
& \begin{tabular}[c]{@{}l@{}}MD=-0.03\\ t=-.01\\ p=.99\\ $\eta_p^2$=.00\end{tabular}     
& \begin{tabular}[c]{@{}l@{}}MD=-5.76\\ t=-2.42\\ \textbf{p=.02*}\\ $\eta_p^2$=.04\end{tabular} 
& \begin{tabular}[c]{@{}l@{}}MD=5.29\\ t=2.22\\ \textbf{p=.03*}\\ $\eta_p^2$=.04\end{tabular}    
& \begin{tabular}[c]{@{}l@{}}MD=-1.11\\ t=-.33\\ p=.74\\ $\eta_p^2$=.00\end{tabular}
& \begin{tabular}[c]{@{}l@{}}MD=-7.37\\ t=-2.18\\ \textbf{p=.03*}\\ $\eta_p^2$=.03\end{tabular}
& \begin{tabular}[c]{@{}l@{}}MD=-3.16\\ t=-.94\\ p=.35\\ $\eta_p^2$=.00\end{tabular}  
& \begin{tabular}[c]{@{}l@{}}MD=-6.26\\ t=-1.86\\ p=.07\\ $\eta_p^2$=.03\end{tabular}
& \begin{tabular}[c]{@{}l@{}}MD=-2.05\\ t=-.61\\ p=.54\\ $\eta_p^2$=.00\end{tabular}
& \begin{tabular}[c]{@{}l@{}}MD=4.21\\ t=1.25\\ p=.21\\ $\eta_p^2$=.01\end{tabular} \\ \hline
\end{tabular}
}
\end{table*}

\section{Discussion}

To answer the five research questions and the corresponding hypotheses about the effect of different variables, we found that each of the factors played a specific role, with \textit{Question Type} having the greatest impact, followed by \textit{Interview Attitude}, \textit{Preparation} and \textit{Realism} all having approximately the same effect, and finally \textit{Timekeeping} having the smallest effect. Our findings both negate and support some of the hypotheses. Our results showed that professional questions, being unprepared, timed answers, and negative interviewers can indeed cause more anxiety than their respective opposites. However, in terms of \textit{Realism}, it was predicted that a greater level of realism would result in greater anxiety, but this turned out not to be the case, where \textit{Realistic VR} was found to have the greatest anxiety-inducing effect.

Regarding the independent variables, we found that \textit{Question type} has the most significant effect among other independent variables. In particular, professional questions can lead to higher anxiety on each dependent variable and dimension. For example, considering anxiety, professional questions can lead to more self-perceived, SCR-embodied, and interviewer-rated anxiety; for overall experience, professional questions can cause more discomfort, more cognitive load, and less eye contact; and for interview performance, professional questions can cause decreased communication skills. The only exception is the overall performance on which \textit{Question Type} has no significant impact. There has been little research on the impact of question types on interview anxiety; also, the impact of many question variables remains poorly understood, for example, question variables including whether the question is open or closed \cite{gee1999colour}, experience-based or situational questions \cite{ellis2002use}, `lower-order’ or `higher-order’ thinking \cite{bradley2008ask}.  Our research focused on professional and personal questions related to job interviews. Our data suggest that professional questions can cause more cognitive load than personal questions. Susan Gee et al. \cite{gee1999colour} mentioned that a recall question requires more cognitive processing than an answer to a recognition question, which offered valuable insight into our study. The professional questions in our survey required a memory search and thus can be defined as recall questions. In contrast, personal questions with specific cues provided were very familiar with recognition questions. However, Susan Gee used a sample of 157 children aged nine to thirteen, and our experiments mainly targeted college students.
Next, \textit{Interview Attitude}, \textit{Preparation}, and \textit{Realism} all have considerable effects on each dependent variable. Apparently, a negative interviewer can cause more SCR-embodied anxiety, interviewer-rated anxiety, less eye contact, and worse performances. Joung Huem Kwon's finding considered that anxiety level was affected more by the attitude of the virtual interviewer than its level of realism \cite{kwon2009study}, whereas our findings do not support that the attitude's impact necessarily outweighed the level of realism. 
Joung Huem Kwon's experiment only focused on virtual humans and did not include a real human interviewer. Also, the indicator of the interviewee's anxiety used in Joung Huem Kwon's study was simply physiological measurements(i.e., the percent rate of gaze fixation and eye blink). 
In a similar study, Patrick Gebhard \cite{gebhard2014exploring} designed two types of virtual recruiters: a sympathetic one with friendly facial expressions and a warm tone; and a demanding one with unfriendly facial expressions and a cold tone. They found that the participants felt that the demanding character induced a higher stress level than the understanding character and felt less comfortable, which is in line with our discovery in terms of the effect on the overall experience.
Similarly, no preparation before an interview can lead to more self-perceived anxiety, more discomfort, a higher cognitive load regarding frustration, and worse performance. This is consistent with a previous study which suggested that job-seekers perform better in job interviews when they are better prepared and have rehearsed answers to common interview questions, and the experiential practice of mock interviews may enhance students’ preparation for real-world job interviewing \cite{hansen2006employment}.
The influence of \textit{Realism} is much more complicated since this variable has four levels, mixed-effect model indicates that \textit{Realism} has a significant impact on self-perceived anxiety, discomfort, cognitive load, eye contact, and communication skills, with further post-hoc analysis, we discovered that \textit{Realistic VR} induces more self-perceived anxiety, more discomfort, higher cognitive load regarding frustration, and less eye contact than \textit{PC} and even \textit{Real Person}, which is against our prior hypotheses that \textit{Real Person} should have caused more anxiety than \textit{Realistic VR}, yet, it is reasonable and in line with many previous findings that VR is effective in inducing stress \cite{wallergaard2011virtual,zimmer2019same,fallon2021multi}.
Also, there is no significant difference shown in any dependent variables between \textit{Realistic VR} and \textit{Cartoon VR} except for \textit{Realistic VR} can reduce eye contact compared to \textit{Cartoon VR}, which aligns with Jean-Luc Lugrin's finding that graphical details or level of realism for avatar visual display reveal no significant differences~\cite{lugrin2015influence}.
Lastly and unexpectedly, \textit{Timekeeping} has the least impact, only shown in interviewer-rated anxiety and performance; specifically, keeping time increases interviewer-rated anxiety and cognitive load considering physical and temporal demand, also leads to worse performance. Nevertheless, the ability to finish tasks under time urgency is crucial; thus, a previous study has validated a Virtual Training System for improving time-limited decision skills and learning performances \cite{doi:10.1089/109493101300117947}, while our research mainly focused on interview performance instead of learning performance as the previous study, both studies indicated the potential of virtual reality as a training tool.

Regarding the dependent variables, results indicate that \textbf{Anxiety} is greatly influenced by \textit{Question type}, secondly \textit{Interviewer attitude}, and lastly \textit{Timekeeping}, \textit{Preparation} and \textit{Realism}; while \textbf{Overall experience} is greatly influenced by \textit{Question type}, \textit{Preparation} and \textit{Realism}, secondly \textit{Interviewer attitude} and \textit{Timekeeping}; yet \textbf{Performance} is effected by all five variables and with almost the same level of influence with \textit{Preparation} slightly having more impact. 
We further investigated the association between dependent variables and found consistent associations between self-perceived anxiety, SCR-embodied anxiety, and interviewer-rated anxiety, especially in \textit{Question Type} and \textit{Realism} where \textit{Realistic VR} tends to induce more anxiety than \textit{PC} and \textit{Real Person}. However, we found an inconsistency between self-perceived performance collected in NASA-TLX and interviewer-rated performance; interviewees tend to believe their performances are influenced by \textit{Question type}, \textit{Preparation} and \textit{Realism} while interviewers think that their performances are mainly affected by \textit{Interviewer attitude}, \textit{Timekeeping} and \textit{Preparation} even though both sides found \textit{Preparation} influenced performances.
The inconsistency might be because the interviewer had a full-body avatar. However, the interviewee only had both hands as a physical presence in virtual reality, where the interviewers could only judge the interviewee's voice without facial expressions or eye contact to rate their performances. Therefore, the interviewer's evaluation might need to be completed in virtual reality. For the interviewer to evaluate the interviewee's performance more comprehensively, the interviewees could also interact with more expressive avatars, such as a customized avatar with facial and motion capture that can deliver their feelings, facial expressions, and body movements in real-time. The previous study also showed that facial animation could increase the enfacement illusion and avatar self-identification \cite{gonzalez2020using}.

Our findings have a few implications for the optimization and development of VRIS: 
(1) Professional questions and an interviewer with a negative attitude can remarkably induce anxiety during an interview; (2) VR interviews can indeed be effectively used to produce a similar interview experience, inducing the same or even  more anxiety and discomfort than real person interviews to the interviewees; (3) low-fidelity avatars can provide the same user experience, anxiety level, cognitive load as the high ones while having lower requirements for computational performance, time latency, network load, and hardware; (4) preparation is still the critical element to have good performance; (5) during an interview, self-perceived anxiety and the interviewer's evaluated anxiety are approximately the same which means that the interviewer can detect the interviewee's tension level well

\section{Limitation}

The quantification of anxiety is tricky, and electrodermal activity responses may not accurately capture the transient nature of anxiety and are influenced by irrelevant factors, including food and drink intake. Thus for future work, we consider using eye movements, facial expressions, voice intonation, physical motions, or nerve center activity to quantify anxiety comprehensively.  
In addition, the influence of long-term studies is not reflected in our short-term, sequential experiments; a long-term consecutive interview study may reveal additional implications for designing VRIS.
Moreover, we can further investigate the relationships between dependent variables to identify whether the higher the anxiety level, the worse the experience or interview performance must be.
\section{Conclusion}
We developed and evaluated a virtual interview simulator to investigate the possible causes of anxiety in job interviews within VRIS. 
By conducting an orthogonal experimental design with eight job interview conditions and evaluating it with 19 college students to assess the significance of 
five possible anxiety-inducing factors, our study sheds light on understanding the fundamental factors creating people's anxiety and influencing experience and performance during an interview. 
Results confirm the significance of specific variables and emphasize the need to consider question types in VRIS.
We also identified the effectiveness of the VR interviews regarding anxiety-inducing and overall experience compared to real-person interviews; therefore, VRIS could be a promising tool in training and practicing for interviews.

\noindent
%If any of the sections are not relevant to your manuscript, please include the heading and write `Not applicable' for that section. 


\section{Appendix}
\subsection{Interview questions and questionnaires}
% An appendix contains supplementary information that is not an essential part of the text itself but which may be helpful in providing a more comprehensive understanding of the research problem or it is information that is too cumbersome to be included in the body of the paper
Our supplementary materials include eight sets of interview questions (i.e., 4 sets of personal questions and 4 sets of professional questions), a ``Self-assessment Questionnaire" and a ``Performance Rating Scale" see \footnote{\href{https://drive.google.com/file/d/1w9UOSQpRuYNbXyoKf2Fi-e2MpY4xlYxW/view?usp=share_link}{\textcolor{blue}{Click to view the online supplementary materials}}}.




% The authors wish to thank A, B, and C. This work was supported in part by
% a grant from XYZ.}

%\bibliographystyle{abbrv}
\bibliographystyle{abbrv-doi}

\bibliography{reference}

\end{document}